\documentclass[prb, superscriptaddress]{revtex4}
\usepackage[normalem]{ulem}
\usepackage{epsfig}
\usepackage{color}
\usepackage{amsmath}
\usepackage{enumitem}
\usepackage{hyperref}
\hypersetup{
	colorlinks = true,
	linkcolor = blue,
	citecolor = blue
}

\newcommand{\be}{\begin{equation}}
\newcommand{\ee}{\end{equation}}

\newcommand{\bea}{\begin{eqnarray}}
\newcommand{\eea}{\end{eqnarray}}

\newcommand{\Li}{{\rm Li}}
\newcommand{\p}{\partial}

\newcommand{\la}{\left\langle}
\newcommand{\ra}{\right\rangle}

\newcommand{\tr}{{\rm \, tr\,}}
\renewcommand{\Re}{{\rm \, Re\,}}

\renewcommand{\vec}[1]{{\bf #1}}
\renewcommand{\phi}{\varphi}
\renewcommand{\epsilon}{\varepsilon}
\def\nn{\nonumber\\}

\renewcommand{\cite}[1]{[\onlinecite{#1}]}

\begin{document}

\title{RKKY interaction in a doped pseudospin-1 fermion system at finite temperature}
\date{\today}

\author{D. O. Oriekhov}
\affiliation{Department of Physics, Taras Shevchenko National University of Kyiv, Kyiv 03680, Ukraine}

\author{V. P. Gusynin}
\affiliation{Bogolyubov Institute for Theoretical Physics, Kyiv, 03680, Ukraine}

\begin{abstract}	
We study the RKKY interaction of magnetic impurities in the $\alpha-\mathcal{T}_3$ model which hosts pseudospin-1 fermions with 
two dispersive and one flat bands. By using the effective low-energy Hamiltonian we calculate the RKKY coupling for impurities placed on the same or 
different sublattices. We find that there are three types of interaction, which depend on the model parameter defining the relative strength of hoppings 
between sublattices, two of them can be reduced to graphene case while the third one is new and is due to the presence of a flat zero-energy band.
We derive general analytical expressions for the RKKY interaction in terms of Mellin-Barnes type integrals and analyze different limiting cases. The cases 
of finite chemical potential and temperature, as well as asymptotic at large distances are considered. We show that the interaction between
impurities located at different rim sites displays a very strong temperature dependence at small doping being a direct consequence of the flat band.
The subtleties of the theorem for signs of the RKKY interaction at zero doping, as applied to the $\mathcal{T}_3$ lattice, related to
the existence of a dispersionless flat band are discussed.
\end{abstract}
\maketitle

\section{Introduction}

The Ruderman-Kittel-Kasuya-Yosida (RKKY) interaction \cite{Ruderman-Kittel} is an indirect exchange interaction between two localized magnetic moments mediated by
a background of electrons. It is an important characteristic of electron system and a fundamental interaction responsible for magnetic ordering in spin
glasses and alloys. Besides three dimensions, it has been extensively studied for the electron gas in one \cite{Yafet} and two \cite{Fischer} dimensions. After
the experimental discovery of graphene, the RKKY interaction in systems with Dirac-like dispersion attracted a great interest \cite{BreySarma2007, saremi2007, Kogan-doped-2012, sherafati-doped-2011,roslyak2013, Cao2019, Kogan2019, Kogan2011, Black2010, sherafati2011} due to the richness of their structures. Moreover,
the final results for the complete structure of the RKKY interaction in graphene were obtained only after a decade of debates \cite{sherafati-doped-2011,klier2015}. The RKKY interaction was studied also in strained graphene \cite{Gorman}, bilayer graphene \cite{Parhizgar2013,Klier2014}, biased single-layer silicene \cite{Zera2019}, 8-Pmmn borophene \cite{Paul2019}, on the surface of three-dimensional Dirac semimetals \cite{Kaladzhyan2019}.

Graphene has given a start to a proliferation of fermionic quasiparticles emerging in condensed matter systems which have no counterparts in
particle physics where Poincar$\acute{e}$ symmetry constrains  fermions to the three types: Dirac, Weyl, and Majorana (not discovered yet) particles with
spin $1/2$. In condensed matter systems, symmetries are less restrictive and besides fermions with pseudospin $1/2$ other types of fermions with a higher pseudospin can appear in two- and three-dimensional solids. A recent paper [\onlinecite{Bradlyn}] has given a classification of possible
low-energy fermionic excitations protected by space group symmetries of lattices in solid state systems with spin-orbit coupling and time-reversal symmetry.
The ${\cal T}_3$ lattice provides one of the well-known realizations of pseudospin-1 fermions in two dimensions \cite{Sutherland,Bercioux}.
Pseudospin-$1$ fermions appear also in the  Lieb \cite{Shen2010} and kagome lattices \cite{Green2010}. Recently an experimental evidence of Dirac
	fermions as well as flat bands was reported in the antiferromagnetic kagome metal FeSn \cite{Kang2019}. Also, the realizations of Lieb lattice as electronic lattice formed by the surface state electrons of Cu(111) \cite{Slot2017Nature} as well as the Lieb-like lattices in covalent-organic frameworks were reported \cite{Jiang2019Nature,Cui2020Nature}.  Fermions of different pseudospins may coexist
in some lattices, for example, Dirac and pseudospin-1 fermions are found to coexist in the $\alpha-\mathcal{T}_3$ model \cite{Malcolm}, the edge-centered
honeycomb lattice \cite{Lan2012}, and the 2D triangular kagome lattice \cite{Wang2018}, Weyl fermions coexist with pseudospin-1 and pseudospin-$3/2$ fermions in transition metal silicides \cite{Tang2017} under the protection of crystalline symmetries.


In this work we analyze the RKKY interaction in the so-called $\alpha-\mathcal{T}_3$ model  \cite{Raoux} which contains the mixing of Dirac and pseudospin-1 fermions as low-energy excitations. The $\alpha-\mathcal{T}_3$  model is a tight-binding model of two-dimensional fermions on the $\mathcal{T}_3$ (or dice) lattice whose atoms are situated at vertices of hexagonal lattice and the hexagons centers \cite{Sutherland,Vidal}.  The parameter $\alpha$
 describes the relative strength of couplings between the honeycomb lattice sites and the central site. Thus, as $\alpha$ changes the $\alpha-\mathcal{T}_{3}$ model reveals a smooth transition from graphene ($\alpha=0$) to dice or $\mathcal{T}_{3}$ lattice ($\alpha=1$). Since
 the $\alpha-\mathcal{T}_3$ model has three sites per unit cell, the electron states in this model are described by three-component fermions. It is natural then
 that the spectrum of the model is comprised of three bands. Two of them form Dirac cones as in graphene, and the third band is completely flat, dispersionless,
 and has zero energy in the whole Brillouin zone \cite{Raoux}. All three bands meet at the $K$ and $K^{\prime}$ points, which are situated at the corners of the Brillouin zone. In the linear order in momentum deviations from the $K$ and $K'$ points, the low-energy Hamiltonian of the dice model with $\alpha=1$
describes massless pseudospin-1 fermions and is given by the scalar product of momentum and the spin-1 matrices.

The $\mathcal{T}_3$ lattice was experimentally realized in Josephson arrays \cite{Abilio1999, Serret} as well as in a network made of metallic wires tailored in a high mobility two-dimensional electron gas \cite{Naud2001}, and its optical realization by laser beams was proposed in Ref.\cite{Rizzi}. The experiments \cite{Abilio1999, Serret, Naud2001} have confirmed the existence of novel localization effects, which arise due to the presence of flat band in the spectrum of $\mathcal{T}_3$ lattice.
Recently several physical quantities have been studied in the $\alpha-{\cal T}_3$ model such as orbital susceptibility \cite{Raoux}, optical and magneto-optical conductivity \cite{Malcolm_2014, Carbotte,Illes,Cserti}, magnetotransport \cite{Malcolm,Biswas,Xu,Islam}. The role of transverse magnetic field on \textit{zitterbewegung} was studied in Ref.\cite{Biswas2018} and the enhancement of thermoelectric properties of a nanoribbon made of $\alpha-\mathcal{T}_3$ model was discussed in a recent paper \cite{Firoz_Islam}. The stability of flat band with respect to different perturbations such as terminations of the lattice as well as the phenomenon of atomic collapse the Coulomb field of the charged impurity were studied in Refs.\cite{Oriekhov2018LTP, Coulomb_alphaT3, Bugajko2019}.

 The presence of completely flat energy band is a remarkable feature of the considered model, for example, it results in strong paramagnetic response in a magnetic field \cite{Raoux}. In general, the Fermi systems hosting flat bands attract a lot of attention last time because quenching of the kinetic energy strongly enhances the role of electron-electron and other interactions and may lead to the realization of many very interesting correlated states. The most striking recent example is the observation of superconductivity in twisted bilayer graphene \cite{Cao-twisted} when tuned to special "magic angles" at which isolated and relatively flat bands appear. The three-bands models with a flat band found their applicability in many physical systems (see, for example, reviews \cite{Khodel2017,Leykam2018}), surprisingly even for the description of  equatorial waves \cite{Delplace2017}.
The special role of flat zero Landau level on RKKY interaction in graphene was analyzed in Ref.\cite{Cao2019}.

 The RKKY interaction of impurities placed on dice lattice demonstrates larger richness compared to graphene. As in case of graphene, the RKKY interaction can be written as a product of oscillating part $f_{ab}(\mathbf{R})$ resulting from intervalley scattering times an interaction integral $I(R)$ ($a,b$ refer to sublattices $A,B,C$). We show that while some relative locations of impurities can be reduced to graphene case (multiplied by $\alpha$ dependent coefficients), there is also a new type of interaction. Like in graphene, the RKKY interaction in undoped $\alpha-\mathcal{T}_3$ model decays as $1/R^3$ while there are envelope oscillations for finite doping at large distances. We also show that in some cases the flat band gives an essential contribution in the RKKY interaction, especially for the undoped case and small temperature.

 The paper is organized as follows: In Sec.\ref{sec:RKKY-general} we discuss a general expression for the RKKY interaction. In Sec.\ref{sec:Alpha-T3-general} we describe the general properties of the $\alpha-\mathcal{T}_{3}$ model and derive the corresponding Green functions in the mixed real space - frequency representation. In Sec.\ref{sec:RKKY-AB} we calculate the RKKY interaction for impurities placed on different sublattices of dice lattice, concentrating on the most interesting case of impurity positions which is absent in graphene. In Appendix \ref{appendix:GF-calculation} we present the expression for the retarded Green's function of pseudospin-1 excitations near $K$ points. In Appendices \ref{sec:integral-chemical-potential} and \ref{appendix:zero-chemical-mu} we derive the exact expressions for interaction integrals in terms of Mellin-Barnes type integrals.
\section{Basic formulas}
\label{sec:RKKY-general}
Generally, the RKKY interaction defined by second-order correction to the free energy $\delta F=\frac{1}{2}T\,\text{Tr} VG_0VG_0$, where trace goes over all degrees of freedom.
Here the free Green function is defined by the standard tight-binding or low energy  Hamiltonian, which contains contributions from both valleys.  The interaction  potential of impurity and electron spins is given by \cite{Kogan2011,Cao2019}
\begin{align}
V^{\left(\mu_{1}, \mu_{2}\right)} \equiv V^{\left(\mu_{1}\right)}+V^{\left(\mu_{2}\right)}=-\lambda\left[\mathbf{S}_{1} \cdot \mathbf{s} \delta\left(\mathbf{r}-\mathbf{R}_{1}\right) P_{\mu_{1}}+\mathbf{S}_{2} \cdot \mathbf{s} \delta\left(\mathbf{r}-\mathbf{R}_{2}\right) P_{\mu_{2}}\right],
\end{align}
where $\mathbf{S}_i$ are the spin operators of  impurities and $\mathbf{s}=\hbar\boldsymbol{\sigma}/2$ is the spin of itinerant electrons. The spin-spin coupling constant can be estimated as $\lambda\simeq 1\mbox{eV}$. The sublattice projectors are denoted by $P_{\mu}$, and can be written as the following diagonal matrices  $P_A=\text{diag}(1,0,0),\quad 	 P_C=\text{diag}(0,1,0)$ and  $P_B=\text{diag}(0,0,1)$.
The contribution, which accounts for the interaction between two different spins, is given by
\begin{align}
&\delta F_{12}=\frac{\lambda^2\hbar^2 }{2} \vec{S}_1 \vec{S}_2  \int\limits_{0}^{1/T} d\tau \,\tr\left[P_{\mu_{1}} G_0(\vec{R}_1,\vec{R}_2;\tau)P_{\mu_{2}} G_0(\vec{R}_2,\vec{R}_1;-\tau)\right].
\label{RKKY-general}
\end{align}
Using the following Fourier decomposition of imaginary-time Green function,
\begin{align}\label{eq:RKKY-graphene-matsubara-fourier}
G_0(\tau)=T \sum_{n} G_0\left(i \omega_{n}\right) e^{-i \omega_{n} \tau},\quad  \omega_{n}=(2 n+1) \pi T,
\end{align}
we can replace the integral over imaginary time $\tau$ by $T \sum\limits_{i\omega_n}$. For example, for $\delta F_{12}$ we get
\begin{align}
\delta F_{12}=\frac{\lambda^2\hbar^2}{2} \vec{S}_1 \vec{S}_2 T \sum_{n}\tr\left[P_{\mu_{1}} G_0(\vec{R}_1,\vec{R}_2;i\omega_n+\mu)P_{\mu_{2}} G_0(\vec{R}_2,\vec{R}_1;i\omega_n+\mu)\right],
\end{align}
where we introduced the chemical potential $\mu$.
Performing the sum over the Matsubara frequencies by means of the formula
\begin{align}
T\sum_{n}f(i\omega_n)=-\int\limits_{-\infty}^\infty\frac{d\omega}{\pi}n_F(\omega){\rm Im}f^R(\omega+i\epsilon),
\label{Matsubara-summation}
\end{align}
where $n_F(\omega)=1/(\exp(\omega/T)+1)$ is the Fermi distribution function and superscript $R$ denotes retarded function.
Hence we find an effective RKKY interaction between two magnetic
impurities with the spins $\mathbf{S}_1$, and $\mathbf{S}_2$, sitting at the positions $\vec{R}_1$ and $\vec{R}_2$
\begin{align}
\delta F_{12}=J_{\mu_1\mu_2}\mathbf{S}_1\mathbf{S}_2,\quad J_{\mu_1\mu_2}=(\lambda^2\hbar^2/4)\chi_{\mu_1\mu_2}(\vec{R}_1,\vec{R}_2),\end{align}
where $\chi$ is the  spin-independent susceptibility, however, it depends upon whether atoms belong to the same or different sublattices.
\begin{align}\label{eq:susceptibility_general}
\chi_{\mu_1\mu_2}(\vec{R}_1,\vec{R}_2)=-\frac{2}{\pi}\int\limits_{-\infty}^\infty d\omega n_F(\omega){\rm Im}\tr\left[P_{\mu_{1}} G_0(\vec{R}_1,\vec{R}_2;\omega+\mu)P_{\mu_{2}} G_0(\vec{R}_2,\vec{R}_1;\omega+\mu)\right].
\end{align}
After calculating the trace, the role of projectors is reduced to taking specific components of Green functions $G_{\mu_1 \mu_2}$ and $G_{\mu_2 \mu_1}$.

\section{Green function of the $\alpha-\mathcal{T}_3$ model}
\label{sec:Alpha-T3-general}

\begin{figure}
	\includegraphics[scale=0.4]{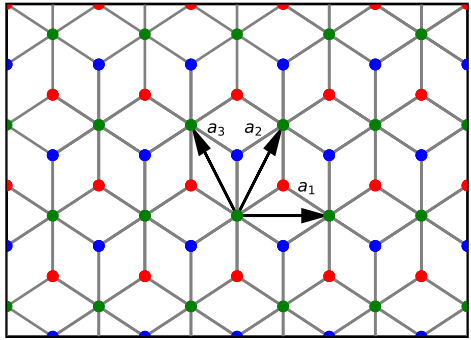}
	\caption{The ${\cal T}_3$ lattice whose red points display atoms of the
		$A$ sublattice, blue points describe the $B$ sublattice, and the green points define the $C$ sublattice. The vectors
		$\vec{a}_1=(\sqrt{3},\,0)d$ and $\vec{a}_2=(\sqrt{3}/2,\,3/2)d$ are the basis vectors of the $C$ sublattice. The nearest neighbor hopping parameters between hub (C) and rim (A, B) atoms are $t_1$ and $t_2$.}
	\label{fig1}
\end{figure}

The $\alpha-\mathcal{T}_3$ model describes quasiparticles in two dimensions with pseudospin $S=1$ on the $\mathcal{T}_3$ lattice schematically shown in
Fig.\ref{fig1}, where $d$ denotes the distance between neighbor atoms. This lattice has a unit cell with three different lattice sites whose
two sites ($A,C$) like in graphene form a honeycomb lattice with hopping amplitude $t_{AC}=t_1$ and additional $B$ sites at the center of each hexagon are
connected to the $C$ sites with hopping amplitude $t_{BC}=t_2$. The C atoms are called hub centers, while A and B are rim sites, and electrons
hop between rim and hub atoms only \cite{Sutherland}. Two hopping parameters $t_1$ and $t_2$ are not equal, in general, and the dice model
corresponds to the limit $t_1=t_2$. The lattice structure and basis vectors are shown on Fig.\ref{fig1}.

We start our description from tight-binding Hamiltonian in two dimensions, which in momentum space reads \cite{Raoux}
\begin{align}
\label{TB-Hamiltonian}
H_{0}(\vec{k})=\left(\begin{array}{ccc}
0 & f_{\vec{k}}\cos\Theta & 0\\
f^{*}_{\vec{k}}\cos\Theta & 0 & f_{\vec{k}}\sin\Theta\\
0 & f^{*}_{\vec{k}}\sin\Theta & 0
\end{array}\right),
\quad \alpha \equiv \tan\Theta=\frac{t_2}{t_1},\quad f_{\vec{k}}=-\sqrt{t_1^2+t_2^2}\,(1+e^{-i\vec{k}\vec{a}_2}+e^{-i\vec{k}\vec{a}_{3}}),
\end{align}
and acts on three-component wave functions with the following order of components $\Psi^T=(\Psi_{A},\Psi_{C},\Psi_{B})$. As was noted in Introduction,
the angle $\Theta$ can be used to interpolate between graphene and dice model. Thus, our results can be  compared with graphene
literature by taking limit $\Theta\to 0$ or $\Theta\to \frac{\pi}{2}$.

The second quantized tight-binding Hamiltonian
	\begin{align}\label{eq:TB-Ham-second-quantized}
		\hat{\mathcal{H}}=\int\limits_{BZ} \frac{d^2 k}{(2\pi)^2} \hat{\Psi}^{\dagger}_{\vec{k}} H_{0}(\vec{k})\hat{\Psi}_{\vec{k}}
	\end{align}
 possesses the particle-hole symmetry, which is realized  by antiunitary operator $\hat{\mathcal{C}}$. It acts on the second quantized wave functions $\hat{\Psi}$ as
\begin{align}\label{eq:symmetry-C}
	\hat{\mathcal{C}} \hat{\Psi} \hat{\mathcal{C}}^{-1}= S\hat{\Psi}^{*},\quad S=\text{diag}(1,\,-1,\,1).
\end{align}
The invariance of the Hamiltonian $\hat{\mathcal{H}}$ under the particle-hole symmetry, $\hat{\mathcal{C}}\hat{\mathcal{H}}\hat{\mathcal{C}}^{-1}=\hat{\mathcal{H}}$, is guaranteed if the following condition is satisfied:
\begin{equation}
SH_{0}(\vec{k})S=-H_0(\vec{k}),
\end{equation}
which is automatically fulfilled for the momentum space Hamiltonian in Eq.\eqref{TB-Hamiltonian}.
Below we show that this symmetry gives restrictions on the sign of the RKKY interactions, similar to the graphene case considered in Ref.\cite{saremi2007}.

It is easy to derive the energy spectrum of the above Hamiltonian, which is qualitatively the same for any $\alpha$ and consists of
three bands: the zero-energy flat band, $\epsilon_0(\mathbf{k})=0$,  whose existence is protected by
 the particle-hole symmetry, and two dispersive bands
\begin{equation}
\epsilon_{\pm}(\mathbf{k})=\pm|f_{k}|=\pm \sqrt{t_1^2+t_2^2}\bigg[3+2(\cos(\vec{a}_1\vec{k})+\cos(\vec{a}_2\vec{k})+
\cos(\vec{a}_3\vec{k}))\bigg]^{1/2}.
\end{equation}
The eigenvectors in the whole Brillouin zone (BZ) are given by Eq.(2) in \cite{Raoux} (gapless case) and by Eq.(5) in \cite{Coulomb_alphaT3}
(gapped case). For dispersionless band the wave function is localized on atoms of sublattices $A,B$ while it is zero on hub atoms $C$.
The presence of a completely flat band with zero energy is perhaps one of the remarkable properties of the $\alpha-\mathcal{T}_{3}$ lattice model.

There are six values of momentum for which $f_{\vec{k}}=0$ and all three bands intersect. They are situated at corners of the hexagonal Brillouin zone. The two inequivalent points, for example, are
\begin{align}
\vec{K}=\frac{2\pi}{d}\left(\frac{\sqrt{3}}{9},\,\frac{1}{3}\right),\quad \vec{K}'=\frac{2\pi}{d}\left(-\frac{\sqrt{3}}{9},\,
\frac{1}{3}\right).
\end{align}
For momenta near the $K$-points, $\vec{k}=\vec{K}(\vec{K}')+\tilde{\vec{k}}$, we find that $f_{\vec{k}}$ is linear in $\tilde{\vec{k}}$,
i.e., $f_{\vec{k}}=\hbar v_F(\xi \tilde{k}_x-i\tilde{k}_y)$ with valley index $\xi=\pm$, where $v_F=3td/2\hbar$ is the Fermi velocity, and in what follows
we omit for the simplicity of notation the tilde over momentum.  As for lattice parameters
 we take their numerical values the same as in graphene.
Hence, in the linear order to momentum deviations from the $K$ and $K'$ points, the low-energy Hamiltonian describes massless pseudospin-1
fermions \cite{Bercioux,Raoux} which for equal hoppings, $\Theta=\pi/4$, is given by the scalar product of momentum and the spin-1 matrices.

\subsection{Green's function}

The lattice Green's function in the tight-binding approximation for Hamiltonian \eqref{TB-Hamiltonian} is given by
\begin{align}\label{eq:RKKY-alphat3-lattice-GF}
G_{0}(\omega,\vec{k})=(\omega-H_0(\vec{k}))^{-1}=\frac{1}{\omega  \left(\omega ^2-|f(\vec{k})|^2\right)} \left(
\begin{array}{ccc}
\omega ^2-\sin^2\Theta \,|f(\vec{k})|^2 & \omega  \cos\Theta f(\vec{k})
& \frac{1}{2} \sin(2\Theta) f(\vec{k})^2 \\
\omega  \cos \Theta f^{*}(\vec{k}) & \omega ^2 & \omega  \sin \Theta  f(\vec{k}) \\
\frac{1}{2}\sin(2\Theta) f^{*}(\vec{k}){}^2 & \omega  \sin\Theta f^{*}(\vec{k}) & \omega ^2-\cos^2\Theta |f(\vec{k})|^2 \\
\end{array}
\right).
\end{align}
In the low-energy model near the $K(K')$ points ($\xi=\pm$), it can be decomposed as
\begin{align}\label{eq:RKKY-alphat3-lowenergy-GF}
G_{0}^{\xi}(\omega,\vec{k})=\frac{1}{\omega  \left(\omega ^2-(\hbar v_F \vec{k})^2\right)} \left(
\begin{array}{ccc}
\omega ^2-\sin^2\Theta \hbar^2 v_F^2 \vec{k}^2 & \omega  \cos\Theta \hbar v_F (\xi k_x-ik_y)
& \frac{1}{2} \sin(2\Theta) (\hbar v_F (\xi k_x-ik_y))^2 \\
\omega  \cos \Theta \hbar v_F (\xi k_x+ik_y) & \omega ^2 & \omega  \sin \Theta  \hbar v_F (\xi k_x-ik_y) \\
\frac{1}{2}\sin(2\Theta) (\hbar v_F (\xi k_x+ik_y))^2 & \omega  \sin\Theta \hbar v_F (\xi k_x+ik_y) & \omega^2-\cos^2\Theta (\hbar v_F \vec{k})^2 \\
\end{array}
\right).
\end{align}
As was shown in Sec.\ref{sec:RKKY-general}, the representation of Green's function in the mixed coordinate-frequency
variables $(\vec{r},\omega)$ is the most useful for the calculation of susceptibility, and related to Eq.\eqref{eq:RKKY-alphat3-lowenergy-GF} by
 Fourier transformation over wave number $\vec{k}$. The Fourier transform of full retarded low-energy Green’s function should contain contributions from both valleys
 \begin{align}\label{eq:GF-valley-decomposition}
 G_{0}(\vec{R}_1, \vec{R}_2, \omega)=\frac{1}{\Omega_{BZ}} \int \frac{d^{2} q}{(2\pi)^2} e^{i \vec{q} \cdot (\vec{R}_1-\vec{R}_2)}\left[e^{i \vec{K}(\vec{R}_1-\vec{R}_2)} G_{0}(\vec{q}+\vec{K}, \omega) +e^{i \vec{K}^{\prime}(\vec{R}_1-\vec{R}_2)} G_{0}\left(\vec{q}+\vec{K}^{\prime}, \omega\right) \right],
 \end{align}
 where $K$ and $K^{\prime}$ are any two adjacent Dirac points in the Brillouin zone, and $\Omega_{BZ}=\frac{2}{3\sqrt{3}d^2}$ is the area of the BZ.
  Replacing wave number by derivative in the matrix part of  \eqref{eq:RKKY-alphat3-lowenergy-GF}, and performing integration as shown in Appendix \ref{appendix:GF-calculation}, we obtain the Green function in valley $\xi$:
 \begin{align}\label{eq:GF-hankel-functions}
 G_{0}^{R}\left(\mathbf{R}_{1}-\mathbf{R}_{2}, \omega,\xi\right)=\frac{1}{\Omega_{BZ}}\frac{\omega}{4(\hbar v_F)^2}\left(\begin{array}{ccc}
 -i\cos^2\Theta H_{0}^{(1)}\left(z\right)& \cos\Theta\xi  e^{-i\xi\phi} H_{1}^{(1)}\left(z\right)& \frac{i}{2}\sin(2\Theta) e^{-2  i\xi\phi} H_{2}^{(1)}\left(z\right)\\
 \cos\Theta\xi e^{i\xi\phi} H_{1}^{(1)}\left(z\right)&-i H_{0}^{(1)}\left(z\right) & \sin\Theta \xi e^{-i\xi\phi} H_{1}^{(1)}\left(z\right)\\
 \frac{i}{2}\sin(2\Theta) e^{2  i\xi\phi} H_{2}^{(1)}\left(z\right) &\sin\Theta \xi e^{i\xi\phi} H_{1}^{(1)}\left(z\right) &-i\sin^2\Theta H_{0}^{(1)}
 \left(z\right)\end{array}\right),
 \end{align}
 where we used notation $z={|\mathbf{R}|(\omega+i\epsilon)}/{\hbar v_{F}}$, and $H_{n}^{(1)}(z)$ is the Hankel function of the first kind. The polar angle of the vector $\vec{R}_{1}-\vec{R}_2$ is denoted by $\phi$. Below we insert Eq.\eqref{eq:GF-hankel-functions} into \eqref{eq:GF-valley-decomposition} and then calculate susceptibility and the RKKY interaction via Eq.\eqref{eq:susceptibility_general} in all 6 relative positions of impurities  AA, AB, BB, AC, BC, CC.
\vskip5mm

\section{RKKY interaction of impurities  on dice lattice}
\label{sec:RKKY-AB}
As was noted before, there are 6 different relative positions of impurities. The corresponding  exchange interactions are
\begin{align}
\label{eq:chi_AA}
&J_{AA}(\vec{R})=\frac{4C
}{\hbar^2 v_F^2}\cos^4\Theta
f_{AA}(\vec{R}) I_0(R,\mu,T),\\
\label{eq:chi_BB}
&J_{BB}(\vec{R})=\frac{4C
}{\hbar^2 v_F^2} \sin^4\Theta
f_{BB}(\vec{R}) I_0(R,\mu,T),\\
\label{eq:chi_CC}
&J_{CC}(\vec{R})=\frac{4C
}{\hbar^2 v_F^2} f_{CC}(\vec{R}) I_0(R,\mu,T),\\
\label{eq:chi_AC}
&J_{AC}(\vec{R})=\frac{4C
}{\hbar^2 v_F^2}\cos^2\Theta
f_{AC}(\vec{R}) I_1(R,\mu,T),\\
\label{eq:chi_BC}
&J_{BC}(\vec{R})=\frac{4C
}{\hbar^2 v_F^2}\sin^2\Theta
f_{BC}(\vec{R}) I_1(R,\mu,T),\\
\label{eq:chi_AB}
&J_{AB}(\vec{R})=\frac{C
}{\hbar^2 v_F^2}\sin^2(2\Theta)
f_{AB}(\vec{R}) I_2(R,\mu,T).
\end{align}
In these expressions we introduced short-hand notations $\vec{R}=\vec{R}_1-\vec{R}_{2}$ and $C={3\lambda^2\hbar^2 d^2}/{64\pi t^2}$. The temperature-independent functions $f_{\mu_1 \mu_2}$ describe oscillations from contribution of different $K$ points for impurities placed on $\mu_1$ and $\mu_2$ sublattices
\begin{align}
&f_{\mu\mu}(\vec{R})=1+\cos
\left(\vec{K}-\vec{K}^{\prime}\right)\vec{R},\\
&f_{AB}(\vec{R})=1+\cos[(\vec{K}-\vec{K}^{\prime})\vec{R}-4 \phi],\quad f_{BA}(\vec{R})=1+\cos[(\vec{K}-\vec{K}^{\prime})\vec{R}+4 \phi],\\ &f_{AC}(\vec{R})=f_{CB}(\vec{R})=1-\cos((\vec{K}-\vec{K}^{\prime})\vec{R}-2\phi),\\ &f_{BC}(\vec{R})=f_{CA}(\vec{R})=1-\cos((\vec{K}-\vec{K}^{\prime})\vec{R}+2\phi).
\end{align}
The functions $f_{\mu_1 \mu_2}$ are the only ones which depend on the direction of the vector $\mathbf{R}$ while other functions are direction-independent.
 In the graphene limit, $\Theta=0$ or $\Theta={\pi}/{2}$,  only three interactions are left, which correspond to coupled lattices  C and A (B). The AB interaction type vanishes in both graphene cases and reaches its maximum value in dice model $\Theta={\pi}/{4}$.

The frequency integrals on the right-hand side of the expressions are
\begin{align}\label{eq:In-definition}
I_{n}(R,\mu,T)=\int_{-\infty}^{\infty} \frac{d \omega f(\omega)}{e^{\frac{\omega-\mu}{T}}+1},\quad f(\omega)=\operatorname{Im}\left[(\omega+i\epsilon)^2 \left(H_{n}^{(1)}\left(\frac{(\omega+i\epsilon) R}{\hbar v_F}\right)\right)^2\right].
\end{align}
We find that the most interesting is the AB case, which cannot be reduced to any known graphene cases due to the lattice geometry, which corresponds to
the appearance of the $H_{2}^{(1)}(z)$ function. For the functions $H_{0}^{(1)}(z+i\epsilon)$ and $H_{1}^{(1)}(z+i\epsilon)$  we can take the limit $\epsilon\to0$ in the integrand, however, this is not the case for $H_{2}^{(1)}(z+i\epsilon)$ due to its more singular behavior when $z \to 0$ which is a reflection of a special role of the flat band with $\omega = 0$. Near $\omega=0$ we find the singular term in the following integral
\begin{align}\label{eq:H2-singular}
(\omega+i\epsilon)^2 \left(H_{2}^{(1)}\left(\frac{ (\omega+i\epsilon) R}{\hbar v_{F}}\right)\right)^{2}\simeq -\frac{16(\hbar v_F)^4}{\pi^2 R^4 (\omega+i\epsilon)^2}-\frac{8(\hbar v_F)^2}{\pi^2 R^2},
\end{align}
hence
\be
{\rm Im}\left[(\omega+i\epsilon)^2 \left(H_{2}^{(1)}\left(\frac{ (\omega+i\epsilon) R}{\hbar v_{F}}\right)\right)^{2}\right]\simeq
\frac{32\epsilon\omega(\hbar v_F)^4}{\pi^2R^4(\omega^2+\epsilon^2)^2}\rightarrow -\frac{16(\hbar v_F)^4}{\pi R^4}\delta'(\omega),\quad \epsilon\to0.
\ee
Adding and subtracting the term ${16(\hbar v_F)^4}/{\pi^2 R^4(\omega+i\epsilon)^2}$ in the expression
\begin{align}
I_{2}(R, \mu, T)= \int_{-\infty}^{\infty} \frac{d \omega }{e^{\frac{\omega-\mu}{T}}+1} \operatorname{Im}\left[(\omega+i\epsilon)^2 \left(H_{2}^{(1)}\left(\frac{(\omega+i\epsilon) R}{\hbar v_{F}}\right)\right)^{2}+\frac{16(\hbar v_F)^4}{\pi^2 R^4(\omega+i\epsilon)^2}-
\frac{16(\hbar v_F)^4}{\pi^2 R^4 (\omega+i\epsilon)^2}\right],
\end{align}
we can safely take the limit $\epsilon=0$ for the first two terms in the square brackets while the third term produces an additional contribution
\begin{align}\label{eq:additional-term-I2}
I_{2}(R, \mu, T)=\int_{-\infty}^{\infty} \frac{d \omega \omega^2  }{e^{\frac{\omega-\mu}{T}}+1} \operatorname{Im}\left[\left(H_{2}^{(1)}
\left(\frac{\omega R}{\hbar v_{F}}\right)\right)^{2}\right]-\frac{4(\hbar v_F)^4}{\pi R^4}\frac{1}{T\cosh^2(\mu/2T)}.
\end{align}
For finite $\mu$ the additional term does not contribute in the zero temperature limit, $T\to0$, while at zero chemical potential, $\mu=0$,
it gives a divergent contribution $\sim-1/T$.

The evaluation of the integral \eqref{eq:In-definition} with $\epsilon=0$ represents a nontrivial task due to the combination of Bessel functions.
It can be written as
\begin{align}\label{eq:In-rewritten}
I_{n}(R,\mu,T)=2\left(\frac{\hbar v_F}{R}\right)^3 \int\limits_{0}^{\infty}dx x^2 J_{n}\left(x\right) Y_n \left(x\right) \left(\frac{1}{z e^{x/a}+1}
+\frac{z}{e^{x/a}+z}-1\right),\quad a=\frac{T R}{\hbar v_F},\quad z=e^{-\mu/T}.
\end{align}
The last term in brackets is divergent at the upper limit, that corresponds to physical divergence at $\omega=-\infty$ in Eq.\eqref{eq:In-definition}.
In such a case one can introduce frequency cut-off, or another well defined regularization \cite{saremi2007,sherafati-doped-2011}. We choose the regularization by replacing $x^2$ by $x^{\alpha-1}$ and  take the limit $\alpha=3$ only in finite expressions. We checked that the frequency cut-off regularization gives the same result. Eq.(\ref{eq:In-rewritten}) is written in terms of the corresponding more general integral $I(\alpha, \nu,z,a)$, Eq.(\ref{general_integral}),  studied
in Appendix \ref{sec:integral-chemical-potential}, as follows
\begin{align}
&I_{n}(R,\mu,T)=\left(\frac{\hbar v_F}{R}\right)^3 I(\alpha=3, n,z,a),\quad n=0,1,\nonumber\\
& I_{2}(R,\mu,T)=\left(\frac{\hbar v_F}{R}\right)^3\left[ I(\alpha=3, n=2,z,a) -\frac{4\hbar v_F}{\pi R T}\frac{1}{\cosh^2(\mu/2T)}\right].
 \label{I_n-corrected}
\end{align}
Generally, the answer can be expressed as inverse Mellin transform (see Eq.\eqref{J-through-Q} or \eqref{Jfinal-withQ}) which is suitable for studying
different physically relevant asymptotics such as low and high temperature expansions, or the behavior at large distances $R$.

\subsection{Small temperature expansion}
To find small temperature corrections at finite chemical potential, one can apply the Sommerfeld expansion for the frequency integral \eqref{eq:In-definition}
rewriting it in the form
\begin{equation}
I_{n}(R,\mu,T)=\int\limits_{-\infty}^\mu d\omega f(\omega)+T\int\limits_0^{\infty}\frac{dx [f(\mu+T x)-f(\mu-T x)}{e^x +1}\simeq \int\limits_{-\infty}^\mu
d\omega f(\omega)+\frac{\pi^2T^2}{6}f^\prime(\mu)+O\left(\frac{T}{\mu}\right)^4.
\label{main-integral-anotherform}
\end{equation}
Using the first equality, one can evaluate interaction numerically. As discussed in Appendix \ref{sec:integral-chemical-potential}, we can find all terms of
the expansion in powers of $T/\mu$. Here we
present only two lowest terms of this expansion, which are given by \eqref{Sommerfeld-J(alpha,nu,u)}.
\begin{align}\label{eq:AB-Sommerfeld}
I_{n}(R,\mu,T)&=\left(\frac{\hbar v_F}{R}\right)^3\left[\frac{1}{\sqrt{\pi}}G^{30}_{24}\left((k_FR)^2\Big|\begin{array}{cc}2,1\\0, \frac{3}{2},
\frac{3}{2}+n,\frac{3}{2}-n\end{array}\right)+\frac{2\pi^{3/2}T^2}{3 \mu^2}
G^{30}_{24}\left((k_FR)^2\Big|\begin{array}{cc} 2,\frac{1}{2}\\ \frac{3}{2}, \frac{3}{2},
\frac{3}{2}+n,\frac{3}{2}-n\end{array}\right)\right],
\end{align}
where we defined the Fermi momentum as $k_F={\mu}/{\hbar v_F}$. Clearly, nonanalytic in the temperature term in $I_2$ (\ref{I_n-corrected}) does not contribute in the Sommerfeld expansion. For zero temperature, using the value of Meijer function at zero argument,
\begin{align}
G^{30}_{24}\left(0\Big|\begin{array}{cc}2,1\\0, \frac{3}{2},\frac{3}{2}+n,\frac{3}{2}-n\end{array}\right)&=\frac{(4n^2-1)\sqrt{\pi}}{8},
\end{align}
we get for exchange integrals of undoped $\alpha-{\cal T}_3$ system
\begin{eqnarray}\label{eq:undoped_interactions}
J^0_{AA}(\mathbf{R})=-\frac{\hbar v_F\cos^4\Theta}{2R^3}C f_{AA}(\mathbf{R}),\quad J^0_{AC}(\mathbf{R})=\frac{3\hbar v_F\cos^2\Theta}{2R^3}C f_{AC}(\mathbf{R}),
\quad  J^0_{AB}(\mathbf{R})=\frac{15\hbar v_F\sin^2(2\Theta)}{8R^3}C f_{AB}(\mathbf{R}).
\end{eqnarray}
For $\Theta=0$, $J^0_{AA}(\mathbf{R})$ and $J^0_{AC}(\mathbf{R})$ coincide with expressions  derived in [\onlinecite{sherafati-doped-2011,klier2015}].
[Note that our definition of the constant $C$ coincides up to a sign with Ref.\cite{klier2015} while Ref.\cite{sherafati-doped-2011}
uses a different definition.] The minus sign for the exchange interaction means ferromagnetic coupling for spins while the positive sign corresponds
to antiferromagnetic one. We see that couplings $J^0_{AB},J^0_{AC}$ describing the interaction of impurities on different sublattices are of antiferromagnetic
nature in undoped $\alpha-{\cal T}_3$ system,  like in the case of graphene \cite{BreySarma2007,saremi2007,sherafati-doped-2011}.
For angles $\Theta$ close to $\pi/4$ (dice model) the coupling $J^0_{AB}$ is significantly larger than graphene-like couplings:
$|J^0_{AB}|>|J^0_{AC}|>|J^0_{AA}|$. All couplings feature $1/R^3$ behavior familiar in graphene.

At finite doping, the short distance (or small $k_F$) behavior is given by
\begin{eqnarray}
&&J_{AA}(\mathbf{R})=J^0_{AA}(\mathbf{R})\left[1-\frac{32(k_FR)^3}{3\pi}\left(\ln\left(\frac{k_FR}{2}\right)+\gamma-\frac{1}{3}\right)\right],\\
&&J_{AC}(\mathbf{R})=J^0_{AC}(\mathbf{R})\left[1-\frac{16(k_FR)^3}{9\pi}\right],\\
&&J_{AB}(\mathbf{R})=J^0_{AB}(\mathbf{R})\left[1-\frac{8(k_FR)^3}{45\pi}\right].
\end{eqnarray}

\begin{figure}
	\centering
\includegraphics[scale=0.65]{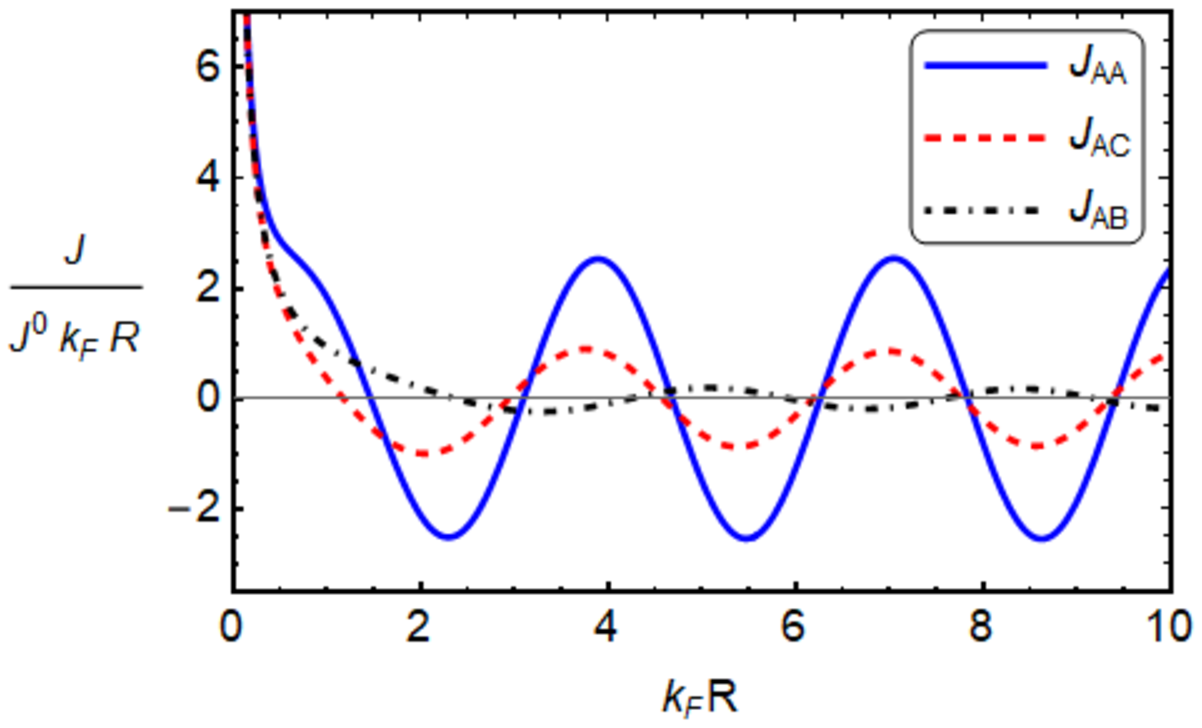}
\includegraphics[scale=0.65]{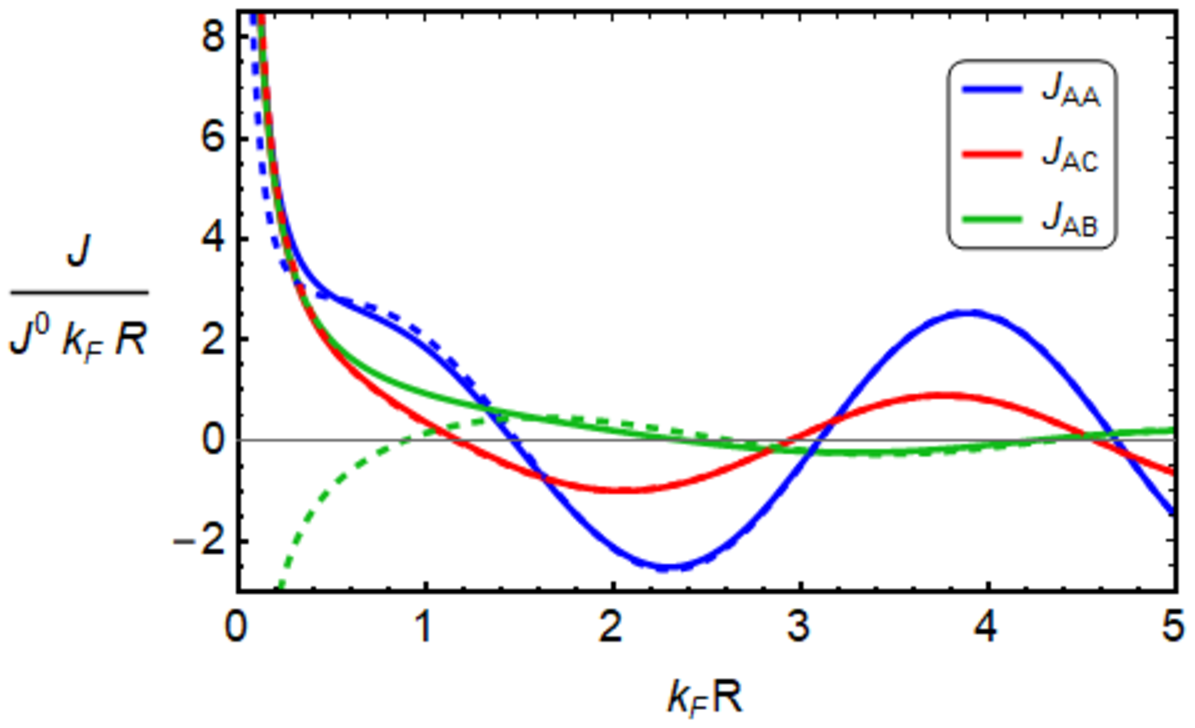}
	\caption{RKKY interactions as functions of $k_FR$ at zero temperature and finite chemical potential calculated through Meijer G-functions. (Left panel) RKKY interactions normalized to their values at $\mu=0$ and divided by $k_FR$. (Right panel) RKKY interactions (solid lines) versus their asymptotic expansions, Eqs.(\ref{eq:sommerfeld-0})-(\ref{eq:sommerfeld-2}),  at $T=0$ (dashed lines) with the same normalizations.}
	\label{fig:zero-T-interactions}
\end{figure}

Expanding Eq.(\ref{eq:AB-Sommerfeld}) at large values $k_F R$, we find the following results for the exchange interactions when both impurities are on the same sublattice $AA$ or couple to different sublattices (AC and AB, for example):
\begin{align}
\label{eq:sommerfeld-0}
&J_{AA}(\vec{R}, \mu,T)=\frac{8}{\pi}J_{AA}^{0}(\vec{R})\left[k_FR\sin(2k_FR)+\frac{1}{4}\cos(2k_FR)-\frac{2\pi^2T^2R^2}{3(\hbar v_F)^2}
\left(k_F R\sin(2k_FR)-\frac{3}{4}\cos(2k_FR)\right)\right],\\
\label{eq:sommerfeld-1}
&J_{AC}(\vec{R}, \mu,T)=\frac{8}{3\pi}J_{AC}^{0}(\vec{R})\left[k_F R \sin \left(2
k_F  R\right)+\frac{5 }{4} \cos
\left(2 k_F R\right)-\frac{2\pi^2  R^2 T^2}{3
	(\hbar v_F)^2 } \left(k_F R
\sin \left(2 k_F R\right)+\frac{1}{4}
\cos \left(2 k_F
R\right)\right)\right],\\
\label{eq:sommerfeld-2}
&J_{AB}(\vec{R}, \mu,T)=-\frac{8}{15\pi}J_{AB}^{0}(\vec{R})\left[k_F R\sin(2k_FR)+\frac{17}{4}\cos(2k_FR)-\frac{2\pi^2T^2R^2}{3(\hbar v_F)^2}
\left(k_FR\sin(2k_FR)+\frac{13}{4}\cos(2k_FR)\right)\right].
\end{align}
One should note that the exchange interactions oscillate with a distance $R$. The terms with $\sin(2k_F R)$ in square brackets are equal
in all cases while more decreasing terms with $\cos{2k_FR}$ are different and have the largest amplitude in case of magnetic impurities situated on sublattices $A$ and $B$. Zero temperature behavior is given by first two oscillating factors in square brackets. A comparison of
Eqs.(\ref{eq:sommerfeld-0})-(\ref{eq:sommerfeld-2}) with the exact formulas (\ref{eq:AB-Sommerfeld}) shows that these asymptotic expressions work
quite well for $k_FR>0.5$ in AA case and $k_F R>1.5$ in AB case (the right panel in Fig.\ref{fig:zero-T-interactions}). We note that while the
normalized couplings $J_{AA}/J^0_{AA},J_{AC}/J^0_{AC}$ oscillate in phase, the coupling $J_{AB}/J^0_{AB}$ oscillates out of phase (see left panel in Fig.\ref{fig:zero-T-interactions}). Physically this is related to the fact that  $A$ atom does not interact directly with
$B$ atom but only indirectly via the hub atom $C$.

\begin{figure}
	\centering
\includegraphics[scale=0.65]{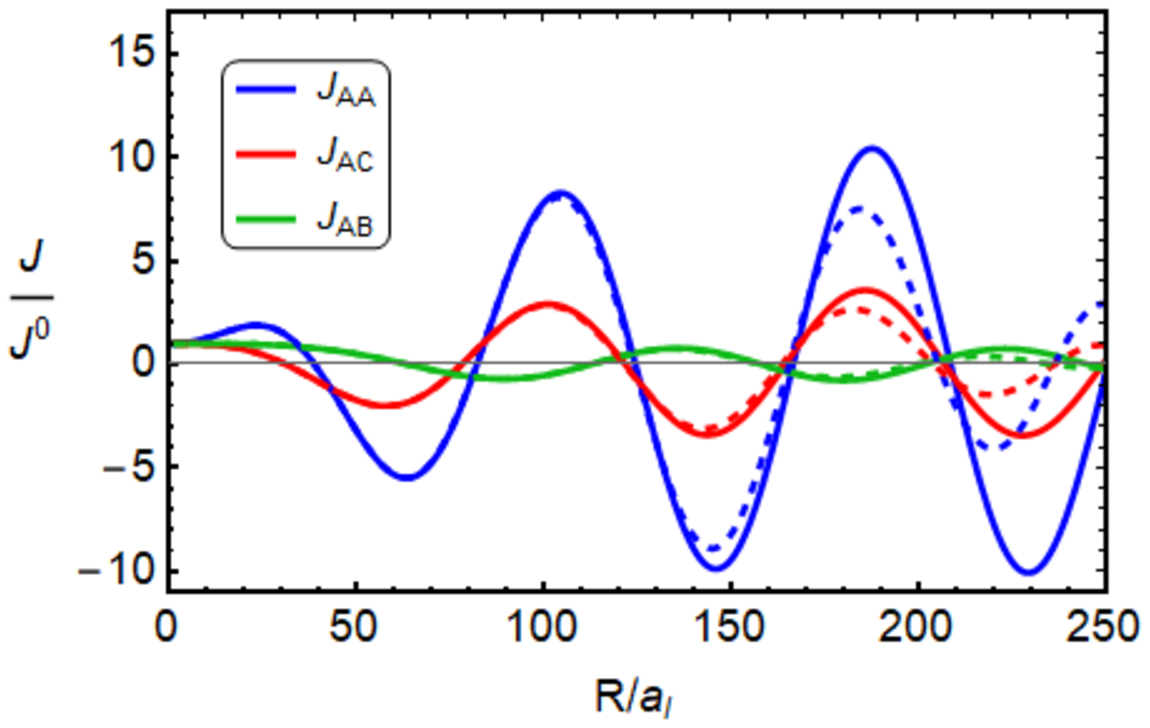}
\includegraphics[scale=0.65]{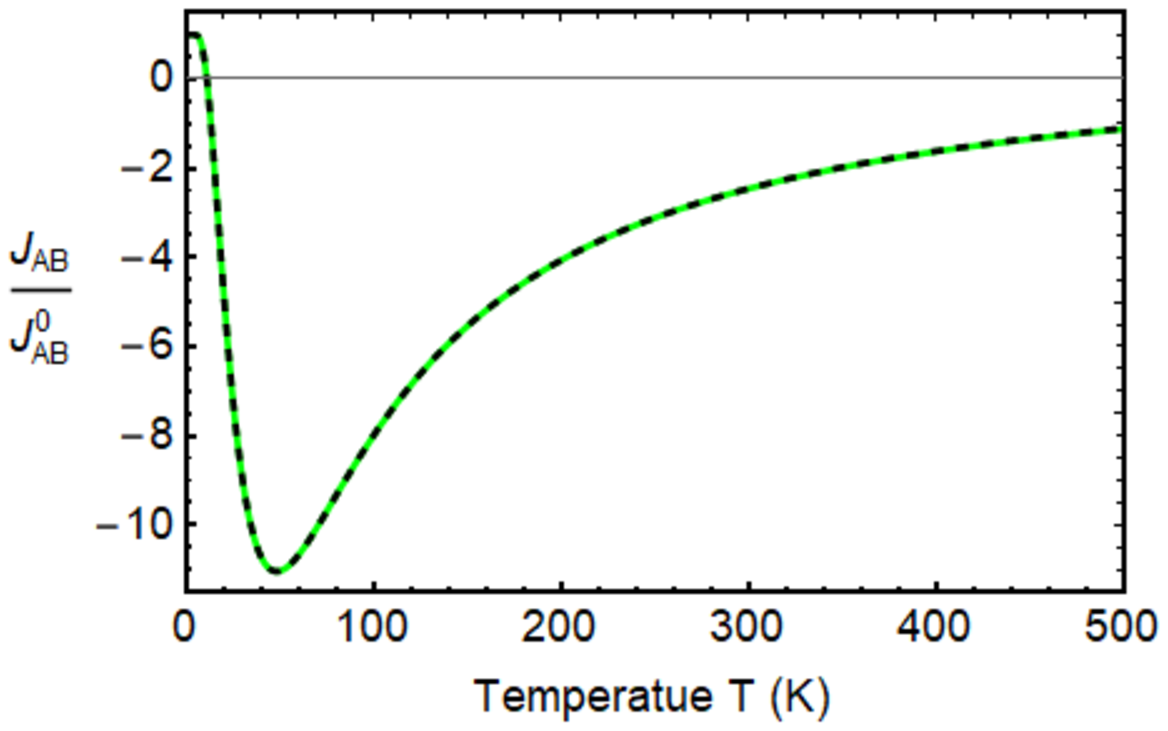}
	\caption{(Left panel) Numerically calculated interactions  (solid lines) are compared with the second-order Sommerfeld expansion \eqref{eq:AB-Sommerfeld} (dashed lines). The chemical potential equals $\mu=0.1\,\text{eV}$ and temperature $T=50\,\text{K}$. Distances are measured in terms of the lattice constant $a_l=\sqrt{3}d=0.246 \,\text{nm}$. The expansion parameter in Eq.\eqref{Q-asymptotical} equals $\frac{2\pi T}{\mu}\approx 0.3$. (Right panel) AB interaction at $R=20 a_l$ and $\mu=0.01\,\text{eV}$ (solid line) and Sommerfeld expansion Eq.\eqref{eq:AB-Sommerfeld} with additional term from Eq.\eqref{eq:additional-term-I2} (dashed line). The nonmonotonic dependence on temperature comes from an additional term in integral \eqref{eq:additional-term-I2}, while the nonsingular part remains constant due to very small value of $(k_F R)^2$. Also we note that the sign of  interaction changes with temperature.}
	\label{fig:exact-finite-T-sommerfeld}
\end{figure}

We also compare the Sommerfeld expansion \eqref{eq:AB-Sommerfeld} with numerically calculated interaction (via the first expression in \eqref{main-integral-anotherform}) at temperature $T=50\,\text{K}$ and chemical potential $\mu=0.1\,\text{eV}$ (see Fig.\ref{fig:exact-finite-T-sommerfeld}). The approximations work very well in a large interval of distances. As one can see from the asymptotic expressions \eqref{eq:sommerfeld-0}-\eqref{eq:sommerfeld-2}, the temperature correction grows with distance. Thus, when $\frac{2\pi^2 T^2 R^2}{3(\hbar v_F)^2}\sim 0.5$,
the next terms in expansion \eqref{Q-asymptotical} become important.

\subsection{Large distance behavior at finite temperature}
In this section we present an exchange interaction in physically relevant case of large distances and finite temperature, thus obtaining more general asymptotic than in Eqs.\eqref{eq:sommerfeld-0}-\eqref{eq:sommerfeld-2}. For this purpose we use the general expansion in powers of ${T}/{\mu}$ (See Eq.\eqref{Sommerfeld-type-expansion} in Appendix \ref{sec:integral-chemical-potential}). However, instead of taking several terms of this expansion we sum up the leading asymptotic terms in series. The obtained Eq.(\ref{I(3,0,z,a)}) allows us to recover approximations similar to those in Ref.\cite{klier2015} using one general expression. Here we present the result for the new $AB$-type interaction integral
\begin{align}\label{eq:klier-expansion-2}
J_{AB}(\vec{R}, \mu, T)=-\frac{8}{15}J_{AB}^{0}(\vec{R})\frac{R}{\hbar v_{F}} F_{1}\left[k_F R \sin \left(2 k_{F} R\right)+\frac{15}{4 } \cos \left(2 k_{F} R\right)+\frac{\pi R}{\hbar v_F}F_{2} \cos \left(2 k_{F} R\right)\right],
\end{align}
where we used the following definitions in analogy with Ref.\cite{klier2015}:
\begin{align}\label{eq:F-definitions-1}
F_{1}=\frac{T}{\sinh \left(\frac{2 \pi T R}{\hbar v_{F}}\right)},\quad F_{2}=\frac{T}{\tanh \left(\frac{2 \pi T R}{\hbar v_{F}}\right)}.
\end{align}
Again in this case the term with $\cos(2k_FR)$ in square brackets has much larger magnitude comparing to the other two interactions $J_{AA}$, $J_{AC}$,
 which are similar to graphene case in \cite{klier2015}. This is an interesting property of $AB$-type interaction.

As was mentioned in Ref.\cite{klier2015}, the term which is proportional to the product $F_1 F_2$ should have a nonmonotonic dependence on temperature. Here we should note that depending on relative distance between impurities, other terms in square brackets in Eq.\eqref{eq:klier-expansion-2} can destroy this effect.

\subsection{Zero chemical potential}
The results in the case of zero chemical potential are not given in the literature in its fullest form even for graphene. Only partial results can be found in the recent paper \cite{Kogan2019}. Here we discuss the asymptotics for low and high temperature which follow from expansion of the expression \eqref{Jfinal-withQ}.

\begin{figure}[h!]
	\centering
	\includegraphics[scale=0.65]{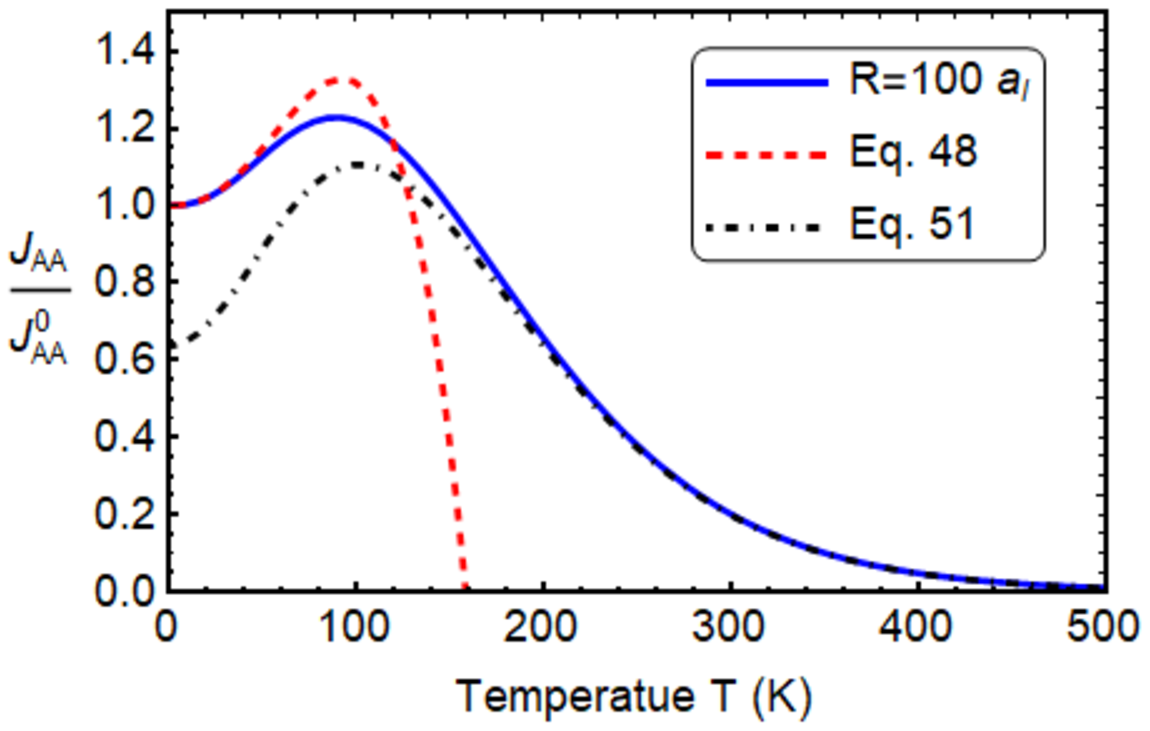}
	\includegraphics[scale=0.65]{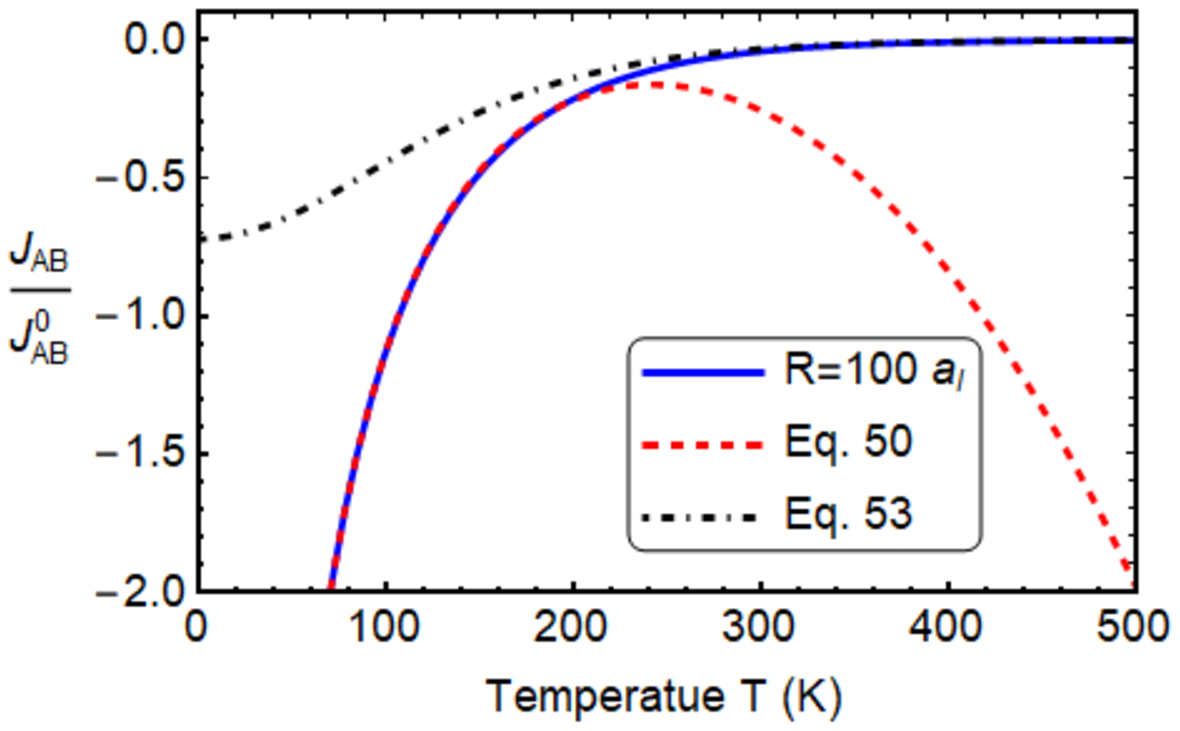}
	\caption{Temperature dependence of normalized interactions $AA$ and $AB$ is compared with asymptotic expressions at small values of parameter $a$ \eqref{eq:small-a-I0},\eqref{eq:small-a-I2} and expansions \eqref{eq:asymp-large-R-0}, \eqref{eq:asymp-large-R-2} at large values of $a$. (Left panel) Non-monotonic behavior of $J_{AA}$ integral, which was discussed in detail in Ref.\cite{klier2015}.  (Right panel) Behavior of relative $AB$ interaction, which has opposite sign comparing to $J_{AB}^{0}$ at zero doping, and becomes very strong as $T$ goes to 0. Such behavior represents a special feature of the $\alpha-T_{3}$ model and is directly related to the existence of flat band.}
	\label{fig:approximations-mu-0}.
\end{figure}

Firstly, we start from the low temperature limit. In fact,  it is easier to determine a low temperature expansion of the integral (\ref{eq:In-definition}) itself.
Making replacement $x\to ax$ in Eq.\eqref{eq:J-definition}, we find
\begin{align}
I_{n}(\mu=0)=\left(\frac{\hbar v_{F}}{R}\right)^{3}\left[-2 C_{2, n}+4 a^3 \int_{0}^{\infty} \frac{x^{2} d x}{e^{x}+1} J_{n}(ax) Y_{n}(ax)\right],
\end{align}
where $a$ is defined in Eq.\eqref{eq:In-rewritten}. Expanding the product of Bessel functions near zero, and then performing integration over $x$, we find the following expressions for interactions:
\begin{align}\label{eq:small-a-I0}
&J_{AA}(\vec{R}, 0,T)=J_{AA}^{0}(\vec{R})\left[1+\frac{16}{\pi } a^3 \left(-6 \zeta(3) \ln (a)-6 \zeta'(3)+\zeta (3) (\ln(16)-9)\right)\right],\\
\label{eq:small-a-I1}
& J_{AC}(\vec{R}, 0,T)=J_{AC}^{0}(\vec{R})\left[1-\frac{16 a^3 \zeta (3)}{\pi}\right],\\
\label{eq:small-a-I2}
& J_{AB}(\vec{R}, 0,T)=J_{AB}^{0}(\vec{R})\left[1-\frac{32}{15\pi a}-\frac{8 a^3 \zeta (3)}{5 \pi}\right],
\end{align}
where $\zeta(x)$ denotes the Riemann zeta-function. Note that the leading temperature correction is of order $T^3$ (or $T^3\log T$) instead of $T^2$ in the case of finite chemical potential (see left panel in Fig.\ref{fig:approximations-mu-0}). In addition one should note the presence of singular $1/T$ term in the $AB$ interaction. As was shown in Eqs.\eqref{eq:H2-singular}-\eqref{eq:additional-term-I2}, this term comes from singular behavior of $H_2$ function, and is related to the effects of flat band. The effect of this term is demonstrated on right panel in Fig.\ref{fig:approximations-mu-0}. Such singular behavior of the AB interaction at low temperature can be used as a benchmark of flat band physics in experiment, for example, in the recently discovered systems \cite{Kang2019,Slot2017Nature}.

The case of high temperatures (or large distances) is much more complicated. The details of calculation are presented in Appendix \ref{appendix:zero-chemical-mu},
and here we present main results for the $AA$, $AC$ and $AB$ cases:
\begin{align}
\label{eq:asymp-large-R-0}
&J_{AA}(\vec{R},0,T)=J_{AA}^{0}(\vec{R})\frac{16 a^2}{\sinh(2\pi a)}\left(\frac{\pi}{\tanh(2\pi a)}-\frac{1}{4a}\right),  \\
\label{eq:asymp-large-R-1}
&J_{AC}(\vec{R},0,T)=J_{AC}^{0}(\vec{R})\frac{ 16a^2}{3\sinh(2\pi a)}\left(\frac{\pi}{\tanh(2\pi a)}+\frac{3}{4a}\right),\\
\label{eq:asymp-large-R-2}
&J_{AB}(\vec{R},0,T)=- J_{AB}^{0}(\vec{R})\frac{16a^2}{15\sinh(2\pi a)}\left(\frac{\pi}{\tanh(2\pi a)}+\frac{15}{4a}\right).
\end{align}
The main difference between the last expression for the $AB$ interaction and the $AA$, $AC$ cases is the changed sign of interaction in Eq.(\ref{eq:asymp-large-R-2}) comparing to Eq.(\ref{eq:small-a-I2}). This change comes from the additional term in Eq.\eqref{eq:additional-term-I2},
which is related to existence of flat band, and exactly cancels $1/R^4$ term in integral, see Appendix \ref{appendix:zero-chemical-mu}.
As is seen, all exchange interactions exponentially decrease at large $RT\gg1$ in the absence of doping. Mathematically this comes from the structure
of Mellin-Barnes integral (\ref{integral-mu=0}), for details we refer the reader to Appendix \ref{appendix:zero-chemical-mu}.

\subsection{Sign of interaction at zero chemical potential and temperature}

For completeness it is worth noting the sign difference between $J_{AB}^{0}(\vec{R})$ and the limit $a\to 0$ in Eq.\eqref{eq:small-a-I2} (which is divergent).
For bipartite lattices, the signs of interactions $J^{0}(\vec{R})$ in undoped case and for zero temperature are fixed by general considerations based on particle-hole symmetry, which result in theorem proved in \cite{saremi2007} (and generalized in \cite{Klier2014}).
Here we find that the same arguments with particle-hole symmetry \eqref{eq:symmetry-C} contain subtleties, which do not allow to fix the sign of $J_{AB}^{0}$.

Using the fact that the ground state is particle-hole symmetric, we find the following symmetry restriction for Green's function:
\begin{align}\label{eq:symmetry-G}
	G_0(\vec{R}_1-\vec{R}_2,\tau_1-\tau_2)=\la\hat{\mathcal{C}} \Psi_1(\vec{R}_1,\,\tau_1) \Psi_2^{\dagger}(\vec{R}_2,\,\tau_2) \hat{\mathcal{C}}^{-1}\ra=-SG_0^{T}(\vec{R}_2-\vec{R}_1,\tau_2-\tau_1)S,
\end{align}
where the operator $\hat{\mathcal{C}}$ and the matrix $S$ are defined in Eq.(\ref{eq:symmetry-C}). Substituting this into susceptibility at zero temperature,
we obtain
\begin{align}
\chi_{\mu_{1} \mu_{2}}\left(\mathbf{R}_{1}-\mathbf{R}_{2}\right)=-\int_{0}^{\infty} d \tau \operatorname{tr}\left[P_{\mu_{1}} G_{0}\left(\mathbf{R}_{1}
-\mathbf{R}_{2} ; \tau\right) P_{\mu_{2}} S G_{0}^{T}\left(\mathbf{R}_{1}-\mathbf{R}_{2} ; \tau\right)S\right].
\end{align}
Calculating the trace, we find susceptibility in terms of single elements of $G_0(\vec{r},\tau)$
\begin{align}\label{eq:chi_mu_mu_tau}
&\chi_{\mu\mu}(\vec{r})=-\int d\tau (G_{0})_{\mu\mu}^2(\vec{r},\tau),\,\, \chi_{AB}(\vec{r})=-\int d\tau (G_{0})_{AB}^2(\vec{r},\tau),\,\, \nn
&\chi_{AC}(\vec{r})=\int d\tau (G_{0})_{AC}^2(\vec{r},\tau),\,\,
\chi_{BC}(\vec{r})=\int d\tau (G_{0})_{BC}^2(\vec{r},\tau).
\end{align}
By using the Fourier transformation of Eq.\eqref{eq:RKKY-alphat3-lattice-GF},
\begin{align}\label{eq:Fourier-tau}
G_{0}(\mathbf{r}, \tau)=\int_{-\infty}^{\infty} \frac{d \omega}{2 \pi} \int_{B Z} \frac{d^{2} k}{(2 \pi)^{2}} G_{0}(\mathbf{k}, i \omega) \exp (-i \omega \tau+i \mathbf{k} \mathbf{r}),
\end{align}
one can easily check that the elements of Green's function in imaginary time representation $G_{0\mu_1\mu_2}(\vec{r},\tau)$ are real. Then, \eqref{eq:chi_mu_mu_tau} gives the following signs for interactions at zero temperature and doping:
\begin{align}
\frac{J_{\mu\mu}^{0}}{|J_{\mu\mu}^{0}|}=-1,\quad \frac{J_{AC}^{0}}{|J_{AC}^{0}|}=\frac{J_{BC}^{0}}{|J_{BC}^{0}|}=1,\quad \frac{J_{AB}^{0}}{|J_{AB}^{0}|}=-1.
\end{align}
Clearly, the sign of $J_{AB}^{0}$ does not agree with our result \eqref{eq:undoped_interactions}. However, one should note that this theorem fixes the sign of interaction only if the integrals in \eqref{eq:chi_mu_mu_tau} exist. This is not the case for the elements $G_{0AB}$ and $G_{0BA}$, because the frequency integral in \eqref{eq:Fourier-tau} diverges at the origin. The divergence comes from the pole at $\omega=0$, which is a manifestation of highly-degenerate flat band.  Therefore we cannot fix the sign of such interaction \textit{a priori}, and should find it from the physically relevant limiting cases, $\mu\to0$ or $T\to0$, and the
answer depends on the order of these limits.

\section{Conclusions}

In recent years,  there was an increasing interest to materials which host fermionic excitations with no analogues in high-energy physics \cite{Bradlyn}.
In particular, the so-called pseudospin-1 fermions  provide a platform for studying exotic physical properties such as transport anomalies, topological Lifshitz transitions, as well as dispersionless flat bands which may lead to the realization of many very interesting strongly correlated states.
Quasiparticle excitations with pseudospin one can be realized in many ways, as we discussed in Introduction.

In this paper we provided results for the RKKY interaction of magnetic impurities, placed on sites of $\mathcal{T}_3$ lattice, mediated by a background
of pseudospin-1 fermions. Our calculations are performed mainly in the low-energy linear-band approximation where we managed to obtain general analytical expressions
for the RKKY interactions which are expressed in terms of Mellin-Barnes type integrals for finite chemical potential and temperature. This allowed us to obtain analytically all asymptotics from one expression. The  asymptotic behavior at large distances was analyzed in detail.  In particular, we found, that oscillatory behavior at large distances was controlled by the same two parameters, the distance between $K$-points and Fermi wave vector, as in graphene.

Our results show that there are three types of interaction, two of them (for impurities on hub and rim sites) can be reduced to graphene case while the third one (between impurities on different rim sites) is new. This new type of interaction, which comes as a special feature of $\mathcal{T}_3$ lattice geometry, becomes very strong at small temperatures and doping. Physically this is an effect of the flat band, which results in a singular  behavior of Green's function at $\omega=0$. For bipartite lattices, it is known that the signs of RKKY interactions
at zero temperature and in the absence of doping are fixed by general considerations based on particle-hole symmetry, which result in the theorem proved in \cite{saremi2007} (and generalized in \cite{Klier2014}).
We discussed the subtleties of this theorem, as applied to the $\mathcal{T}_3$ lattice, related to the existence of a dispersionless flat band.
The breakdown of the theorem for the interaction $J_{AB}^{0}$ is refered to the divergence of the Green's function at zero energy due to flat band. The divergence
is regularized in the presence of finite temperature and/or doping, but taking the limits $\mu=0$ and $T=0$ depends on the order of these limits what is reflected
in the last term in the integral $I_2(R,\mu,T)$ of  Eq.(\ref{I_n-corrected}). This dramatic change of behavior could be utilized to reveal the presence of a flat
band in experiment 
and can be tested, for example, in recently discovered flat-band systems, such as kagome metal FeSn \cite{Kang2019}, Lieb-like lattices in covalent-organic frameworks \cite{Jiang2019Nature,Cui2020Nature} or the electronic Lieb lattice formed by the surface state electrons of Cu(111) \cite{Slot2017Nature}. The RKKY interaction may lead to the realization of magnetic order in these materials.

The described strong temperature dependence in $\alpha-\mathcal{T}_3$ lattice systems may manifest also in Friedel oscillations. The last ones could
be detected  using STM-based quasiparticle interference measurements  \cite{Hasan2018Review}.
 As is known, the flat band emerging in tiny-angle twisted bilayer graphene  results in a strong sensitivity to perturbations leading to strongly correlated states  including superconductivity \cite{Cao-twisted}. While the RKKY interaction was already studied in bilayer graphene \cite{klier2015,Klier2014},
the corresponding calculation for twisted bilayer graphene is still ahead.

\acknowledgements
We are grateful to E.V. Gorbar for useful remarks. V.P.G. acknowledges support by the National Academy of Sciences of Ukraine grant ``Functional properties of materials prospective for nanotechnologies'' (project No. 0120U100858) and collaboration with the Ukrainian-Israeli Scientific Research Program of the Ministry of Education and Science of Ukraine (MESU) and
the Ministry of Science and Technology of the state of Israel (MOST).

\appendix
\section{Green's function in coordinate-frequency representation}
\label{appendix:GF-calculation}
The contribution to the retarded Green's function in $r$ space \eqref{eq:GF-valley-decomposition} from one $K$ point is given by Fourier transform
\begin{align}
G_{0}^{R}\left(\mathbf{R}_{1}-\mathbf{R}_{2}, \omega,\xi\right)=\frac{1}{\Omega_{BZ}}\int \frac{d^{2} k}{(2 \pi)^{2}} e^{i \mathbf{k}\left(\mathbf{R}_{1}-\mathbf{R}_{2}\right)} G_{0}^{\xi}(\mathbf{k}, \omega+i \varepsilon).
\end{align}
Using the expression for Green function in the low energy model \eqref{eq:RKKY-alphat3-lowenergy-GF} and replacing wave numbers by derivatives, we write
\begin{align}
&G_{0}^{R}\left(\vec{r}, \omega,\xi\right)=\frac{1}{\omega}\left(
\begin{array}{ccc}
\omega ^2+\sin^2\Theta \hbar^2 v_F^2 \p_{\vec{r}}^2 & -i\omega  \cos\Theta \hbar v_F (\xi \p_x-i\p_y)
& -\frac{1}{2} \sin(2\Theta) (\hbar v_F (\xi \p_x-i\p_y))^2 \\
-i\omega  \cos \Theta \hbar v_F (\xi \p_x+i\p_y) & \omega ^2 & -i\omega  \sin \Theta  \hbar v_F (\xi \p_x-i\p_y) \\
-\frac{1}{2}\sin(2\Theta) (\hbar v_F (\xi \p_x+i\p_y))^2 & -i\omega  \sin\Theta \hbar v_F (\xi \p_x+i\p_y) & \omega^2+\cos^2\Theta (\hbar v_F \p_{\vec{r}})^2 \\
\end{array}
\right)\times\nn
&\times\frac{1}{\Omega_{BZ}}\int \frac{d^{2} k}{(2 \pi)^{2}} \frac{e^{i \mathbf{k}\vec{r}}}{(\omega+i\epsilon)^2-(\hbar v_F \vec{k})^2}.
\end{align}
Now we integrate over the angle and then use the formula 2.12.4.28 from book \cite{PrudnikovII},
\begin{align}
\int_{0}^{\infty} \frac{x^{\nu+1} J_{\nu}(c x)}{x^{2}+z^{2}}  d x=z^\nu K_{\nu}(c z), \quad  c>0,\,\Re z>0,
\end{align}
and get
\begin{align}
F(\mathbf{r})=\int\frac{d^{2}k}{(2\pi)^{2}}
\frac{ e^{i\mathbf{k}{\bf r}}}{(\omega+i\epsilon)^{2}-(\hbar v_{F}\mathbf{k})^{2}}=\int\limits_{0}^{\infty}
\frac{dk k}{2\pi}\frac{J_{0}(k|\mathbf{r}|)}{(\omega+i\epsilon)^{2}-(\hbar v_{F}k)^{2}}
=-\frac{1}{2\pi(\hbar v_{F})^{2}}K_{0}\left(\frac{-i|\mathbf{r}|(\omega+i\epsilon)}{\hbar v_{F}}\right),
\end{align}
where $J_{0} and K_{0}$ are the Bessel's functions. Using the relation between Macdonald's functions and the Hankel function of first kind,
\begin{align}
H_{\nu}^{(1)}(z)=-\frac{2i}{\pi}e^{-\frac{i\pi\nu}{2}}K_{\nu}\left(ze^{-\frac{i\pi}{2}}\right),\quad z=\frac{|\mathbf{r}|(\omega+i\epsilon)}{\hbar v_{F}},
\end{align}
we find
\begin{align}
&F(\vec{r})=-\frac{i}{4(\hbar v_{F})^{2}}H_{0}^{(1)}\left(\frac{|\mathbf{r}|(\omega+i\epsilon)}{\hbar v_{F}}\right).
\end{align}
Next, we evaluate all matrix elements of Green's function. Let's calculate all needed derivatives
\begin{align}
&(\hbar v_F)^2\p_{\vec{r}}^2 F(\vec{r})=
\frac{i\omega^2}{4(\hbar v_{F})^{2}}H_{0}^{(1)}\left(z\right),\\
&\hbar v_F(\xi\p_x\pm i\p_y) F(\vec{r})=\xi\frac{i\omega e^{\pm i\xi\phi}}{4(\hbar v_{F})^{2}} H_{1}^{(1)}\left(z\right),\\
&(\hbar v_F)^2(\xi\p_x\pm i\p_y)^2 F(\vec{r})=-\frac{i\omega^2 e^{\pm 2  i\xi\phi}}{4(\hbar v_{F})^{2}} H_{2}^{(1)}\left(z\right).
\end{align}
Substituting these expressions back to Green's function, we find result which is given by Eq.\eqref{eq:GF-hankel-functions} in the main text. Note that all elements of the Green function are proportional to $\omega$.

\vskip5mm

\section{Evaluation of the interaction integral}
\label{sec:integral-chemical-potential}

In this Appendix we consider the integral
\be
I(\alpha, \nu,z,a)=2\int\limits_{0}^{\infty}dx x^{\alpha-1} J_{\nu}\left(x\right) Y_\nu \left(x\right) \left(\frac{1}{z e^{x / a}+1}+\frac{z}{e^{x / a}+z}-1\right),
\quad -1<{\rm Re}\,\alpha<1.
\label{general_integral}
\ee
In the region $0<\alpha<1$ we can calculate the terms in round brackets separately, for example,
the term with $-1$  can be evaluated using Eq.2.24.3.1 from the book \cite{Prudnikov3},
	\begin{align}
	C_{\alpha,\nu}=\int\limits_0^\infty dz z^{\alpha-1} J_\nu(z)Y_\nu(z)=-\frac{1}{2\sqrt{\pi}}\frac{\Gamma\left(\frac{\alpha}{2}\right)
		\Gamma\left(\frac{\alpha}{2}+\nu\right)}{\Gamma\left(\frac{1+\alpha}{2}\right)\Gamma\left(1+\nu-\frac{\alpha}{2}\right)},
	\label{C-mu-nu-constant}
	\end{align}
	which gives the following values for $\alpha=3$ and $\nu=0,1,2$:
	\begin{align}
	C_{3,0}=\frac{1}{16},\quad  C_{3,1}=-\frac{3}{16},\quad
	C_{3,2}=-\frac{15}{16}.
	\end{align}
Thus, we can rewrite the integral as follows
\begin{align}
I(\alpha, \nu,z,a)=-2C_{\alpha,\nu}+J(\alpha, \nu,z,a),
\end{align}
where, for $\nu\geq0$,
\be\label{eq:J-definition}
J(\alpha, \nu,z,a)=2\int\limits_{0}^{\infty}dx x^{\alpha-1} J_{\nu}\left(x\right) Y_\nu \left(x\right) \left(\frac{1}{z e^{x / a}+1}+\frac{z}{e^{x / a}+z}\right),
\quad {\rm Re}\,\alpha>0.
\ee
We calculate the last integral using the Mellin transform
\begin{align}
J(\alpha, \nu,z,s)=\int_{0}^{\infty} d a a^{s-1} J(\alpha, \nu, z,a)=2\int_{0}^{\infty} d x x^{\alpha-1} J_{\nu}(x) Y_{\nu}(x) \int_{0}^{\infty} d a a^{s-1}\left(\frac{1}{z e^{x / a}+1}+\frac{z}{e^{x / a}+z}\right).
\label{integral=mu-neq0}
\end{align}
After the change $a\to a x$ and then $a\to 1/a$ Eq.(\ref{integral=mu-neq0}) takes the form
\be
J(\alpha, \nu,z, s)=\int_{0}^{\infty} d x x^{\alpha+s-1} J_{\nu}(x) Y_{\nu}(x) Q(s,z),
\quad 0<\alpha+s<1,
\label{Mellin-transformI}
\ee
where
\be
Q(s,z)=2\int_{0}^{\infty} d a a^{-s-1}\left(\frac{1}{z e^{ a}+1}+\frac{z}{e^{a}+z}\right),\quad {\rm Re\,s}<0.
\label{function-Q}
\ee
The function $Q(s,z)$ possesses the symmetry $Q(s,1/z)=Q(s,z)$.
The integral over $x$ in Eq.(\ref{Mellin-transformI}) is evaluated using Eq.(\ref{C-mu-nu-constant}). There exists the range of parameters $\alpha,s$
where the Mellin transform $J(\alpha,\nu,s,z)$ is defined. We obtain
\be
J(\alpha,\nu,s,z)=-\frac{\Gamma\left(\nu+\frac{\alpha+s}{2}\right)\Gamma\left(\frac{\alpha+s}{2}\right)}
{2\sqrt{\pi}\Gamma\left(\frac{\alpha+1+s}{2}\right)\Gamma\left(\nu+1-\frac{\alpha+s}{2}\right)}Q(s,z),\quad 0<\alpha+{\rm Re}\,s<0,\quad \nu \geq 0,
\label{Q-final-finiteTmu}
\ee
hence
\be
I(\alpha, \nu,z,a)=-2C_{\alpha,\nu}-\frac{1}{2\pi i}\int\limits_{\gamma-i\infty}^{\gamma+i\infty} ds\,a^{-s}\frac{\Gamma\left(\nu+\frac{\alpha+s}{2}\right)\Gamma\left(\frac{\alpha+s}{2}\right)}
{2\sqrt{\pi}\Gamma\left(\frac{\alpha+1+s}{2}\right)\Gamma\left(\nu+1-\frac{\alpha+s}{2}\right)}Q(s,z),
\label{J-through-Q}
\ee
where the contour separates poles of the function $Q(s,z)$ (at $s=0$ and $s=2n+1$, n=0,\,1,\dots, see below) from poles of gamma functions in the numerator.
The integrals in Eq.(\ref{function-Q})  can be evaluated explicitly through the polylogarithm function \cite{Whittaker} and we get
\be
Q(s,z)=-2\Gamma(-s)\left[{\rm Li}_{-s}(-1/z)+{\rm Li}_{-s}(-z)\right].
\label{Q-through-polylogs}
\ee
The function ${\rm Li}_s(z)$ has the following properties. It is an analytical function of complex variables $s,z$. For fixed
$z$, it does not have poles or branch cuts in a finite region of complex $s$-plane, the point $s=\infty$ is the only (essential)
singularity. For fixed $s$,  ${\rm Li}_s(z)$ does not have poles and essential singularities but has a cut in the $z$-plane along the
interval $[1,\infty]$, where it is continuous from below side of the cut. It has the symmetry property with respect to complex conjugation
${\rm Li}_{s^*}(z^*)={\rm Li}^*_s(z)$ for $z$ not belonging to the interval $(-\infty,0)$.

Analytic continuation of $\Li_{s}(z)$ into the region ${|z|>1}$ can be performed by means of the formula (see Eq.(1.11.16) in \cite{Bateman1})
\be
{\rm Li}_s(z)+e^{i\pi s}{\rm Li}_s\left(\frac{1}{z}\right)= \frac{(2\pi)^s}{\Gamma(s)}e^{i\pi s/2}\zeta\left(1-s,\frac{1}{2}+\frac{\ln(-z)}{2\pi i}\right),
\quad {\rm Re}\,s<0,
\ee
where $\zeta(s,q)$ is the Hurwitz $\zeta$-function. When $s$ is a negative even integer, $s=-2m$, $m=1,2,\dots$, we get
${\rm Li}_{-m}(-z)+{\rm Li}_{-m}\left(-{1}/{z}\right)=0$. It follows then from Eq.(\ref{Q-through-polylogs}) that $Q(s,z)$ has poles only for $s=0$ and
odd positive $s=2n+1$, $n=0,1,\dots$, while for even positive $s=2n$ the poles of $\Gamma(-s)$ are canceled by zeros of the sum of polylogarithm functions.
Applying this formula to Eq.(\ref{Q-through-polylogs}) we get
\be
Q(s,z)= -\frac{1}{(2\pi)^{s}\cos(\pi s/2)}\left[\zeta\left(1+s,\frac{1}{2}+\frac{\ln z}{2\pi i}\right)+\zeta\left(1+s,\frac{1}{2}-
\frac{\ln z}{2\pi i}\right)\right].
\label{Q-through-zetas}
\ee
Near $s=0$ the function $Q(s,z)$ behaves as
\be
Q(s,z)\simeq -\frac{2}{s},
\ee
then moving the contour in Eq.(\ref{J-through-Q}) to slightly right of the point $s = 0$ ($\gamma > 0$) and calculating the residue at $s = 0$,
we get
\begin{align}
I(\alpha, \nu,z,a)=-\frac{1}{2\pi i}\int\limits_{\gamma-i\infty}^{\gamma+i\infty} ds\,a^{-s}\frac{\Gamma\left(\nu+\frac{\alpha+s}{2}\right)\Gamma\left(\frac{\alpha+s}{2}\right)}
{2\sqrt{\pi}\Gamma\left(\frac{\alpha+1+s}{2}\right)\Gamma\left(\nu+1-\frac{\alpha+s}{2}\right)}Q(s,z)
\label{Jfinal-withQ}
\end{align}
[the residue at $s=0$ cancels the first term in Eq.(\ref{J-through-Q})].

Expanding the functions $\zeta(s,1/2\pm iv)$ (where $v=\frac{\ln z}{2\pi}$) in series around $v=0$,
we find the following representation of the function $Q(s,z)$ near the point $z=1$:
\begin{align}
Q(s,z) =-\frac{2}{(2\pi)^{s}\cos(\pi s/2)}\sum\limits_{k=0 }^\infty\frac{(-1)^k\Gamma(1+s+2k)\zeta(2k+1+s,1/2)}{\Gamma(1+s)(2k)!}
\left(\frac{\ln z}{2\pi}\right)^{2k}.
\label{Q-z-near1}
\end{align}
This expansion can be used to find a high temperature expansion of Eq.(\ref{Jfinal-withQ}), hence the integral \eqref{eq:In-definition}, when $|\mu|/(2\pi T)\ll1$.

To obtain the expansion at large $|v|=|\mu|/(2\pi T)\gg1$  we start from the asymptotic expansion \cite{Drukarev}:
\begin{align}
\zeta(s,q)=\frac{1}{\Gamma(s)}\sum\limits_{k=0}^\infty\frac{\left(2^{1-2k}-1\right)B_{2k}\Gamma(s+2k-1)}{(2k)!(q-1/2)^{s+2k-1}},
\end{align}
where $B_{2k}$ are Bernoulli numbers. For the function $Q(s,z)$ we get the asymptotic series at large $|v|$:
\begin{align}
Q(s,z)=-\frac{2}{(2\pi |v|)^s\Gamma(s+1)}\sum\limits_{k=0}^\infty\frac{(-1)^k\left(2^{1-2k}-1\right)B_{2k}\Gamma(s+2k)}{(2k)!v^{2k}}.
\label{Q-asymptotical}
\end{align}
The first terms of the expansion of $Q(s,z)$ at small $z$ (large $|v|$) are:
\be
Q(s,z)=-\frac{2}{(2\pi |v|)^s\Gamma(s+1)}\left[\Gamma(s)+ \frac{\Gamma(s+2)}{24v^2}+\frac{7\Gamma(s+4)}{5760v^4}+O\left(\frac{1}{v^6}\right)\right].
\label{Q-expansion_large-mu}
\ee
Hence, for small $z$ (or $\mu/T \gg1$) we obtain, keeping two lowest terms,
\begin{align}
I(\alpha, \nu,u)=\frac{1}{2\pi i}\int\limits_{\gamma-i\infty}^{\gamma+i\infty} ds\,u^{-s}\frac{\Gamma\left(\nu+\frac{\alpha+s}{2}\right)\Gamma\left(\frac{\alpha+s}{2}\right)}
{2\sqrt{\pi}\Gamma\left(\frac{\alpha+1+s}{2}\right)\Gamma\left(\nu+1-\frac{\alpha+s}{2}\right)}
\left[\frac{\Gamma\left(\frac{s}{2}\right)}{\Gamma\left(1+\frac{s}{2}\right)}+\frac{1}{6v^2}
\frac{\Gamma\left(1+\frac{s+1}{2}\right)}{\Gamma\left(\frac{s+1}{2}\right)}\right],\quad u=\frac{\mu R}{\hbar v_F}=k_FR.
\end{align}
Changing $s\to 2s$ and calculating integrals we get equivalent expressions
\begin{align}
I(\alpha, \nu,u)&=\frac{1}{\sqrt{\pi}}G^{30}_{24}\left(u^2\Big|\begin{array}{cc}\frac{\alpha+1}{2},1\\0, \frac{\alpha}{2},
	\frac{\alpha}{2}+\nu,\frac{\alpha}{2}-\nu\end{array}\right)+\frac{1}{6\sqrt{\pi}v^2}
G^{30}_{24}\left(u^2\Big|\begin{array}{cc}\frac{\alpha+1}{2},\frac{1}{2}\\ \frac{3}{2}, \frac{\alpha}{2},
	\frac{\alpha}{2}+\nu,\frac{\alpha}{2}-\nu\end{array}\right)\nn
&=-\frac{1}{\sqrt{\pi}}G^{21}_{24}\left(u^2\Big|\begin{array}{cc}1,\frac{\alpha+1}{2}\\ \frac{\alpha}{2},
	\frac{\alpha}{2}+\nu,\frac{\alpha}{2}-\nu,0\end{array}\right)-
\frac{1}{6\sqrt{\pi}v^2}
G^{21}_{24}\left(u^2\Big|\begin{array}{cc}\frac{1}{2},\frac{\alpha+1}{2}\\  \frac{\alpha}{2},
	\frac{\alpha}{2}+\nu,\frac{\alpha}{2}-\nu,\frac{3}{2}\end{array}\right),
\label{Sommerfeld-J(alpha,nu,u)}
\end{align}
where we used Eq.8.2.1.17 from [\onlinecite{Prudnikov3}],
\bea
G^{mn}_{pq}\left(z\Big|\begin{array}{cc}(a_{p-1}),b\pm l\\ b,(b_{q-1})\end{array}\right)=(-1)^lG^{m-1,n+1}_{p,q}
\left(z\Big|\begin{array}{cc}b\pm l,(a_{p-1})\\ (b_{q-1}),b\end{array}\right).
\eea
The first term in Eq.(\ref{Sommerfeld-J(alpha,nu,u)}) corresponds to the case of zero temperature, and for $\alpha=3$, $\nu=0,1$ it agrees with
the result of Ref.[\onlinecite{sherafati-doped-2011}].
In general, the expansion of the expression (\ref{Q-asymptotical}) over $1/|v|$ corresponds to the expansion over $T/\mu$ (Sommerfeld's expansion).
At large $k_FR$, Eq.(\ref{Sommerfeld-J(alpha,nu,u)}) gives for interested cases $\alpha=3,\nu=0,\,1,\,2$ the results in Eqs. \eqref{eq:sommerfeld-0}-\eqref{eq:sommerfeld-2}.

From our final formula (\ref{Jfinal-withQ}) we can obtain an expansion for $\mu$ near zero by means of Eq.(\ref{Q-z-near1}), and
an expansion for $T\ll\mu$ using Eq.(\ref{Q-asymptotical}).

To find a large $k_F R$ expansion at fixed $RT/\hbar v_F$ we consider the expression (\ref{Jfinal-withQ}) using $Q(s,z)$
represented by the asymptotic series (\ref{Q-asymptotical}),
\bea
I(\alpha, \nu,z,a)&=&\frac{1}{\sqrt{\pi}}\sum\limits_{k=0}^\infty\frac{(-1)^k\left(1-2^{2k-1}\right)B_{2k}}{(2k)!v^{2k}}\frac{1}{2\pi i}
\int\limits_{\gamma-i\infty}^{\gamma+i\infty}\hspace{-3mm} ds\,(2\pi a v)^{-s}\frac{\Gamma\left(\nu+\frac{\alpha+s}{2}\right)\Gamma\left(\frac{\alpha+s}{2}\right)
	\Gamma\left(k+\frac{s}{2}\right)\Gamma\left(k+\frac{1+s}{2}\right)}{\Gamma\left(\frac{1+s}{2}\right)\Gamma\left(1+\frac{s}{2}\right)\Gamma\left(
	\frac{1+\alpha+s}{2}\right)\Gamma\left(1+\nu-\frac{\alpha+s}{2}\right)}\nonumber\\
&=&\frac{2}{\sqrt{\pi}}\sum\limits_{k=0}^\infty\frac{(-1)^k\left(1-2^{2k-1}\right)B_{2k}}{(2k)!v^{2k}}G^{40}_{35}\left((2\pi a v)^2\Big|
\begin{array}{c}\frac{1}{2},1,\frac{1+\alpha}{2}\\ k,k+\frac{1}{2},\frac{\alpha}{2},\frac{\alpha}{2}+\nu,\frac{\alpha}{2}-\nu\end{array}\right),
\label{Sommerfeld-type-expansion}
\eea
where we used the duplication formula for $\Gamma(2k+s)$ and $\Gamma(1+s)$. Since $2\pi av=k_FR$, we consider the asymptotic of Meijer function at large
$k_FR\gg1$. For $\alpha=3$ and nonnegative integer $\nu$ we get
\be
G^{40}_{35}\left((2\pi a v)^2\Big|\begin{array}{c}\frac{1}{2},1,2\\ k,k+\frac{1}{2},\frac{3}{2},\frac{3}{2}+\nu,\frac{3}{2}-\nu\end{array}\right)
\simeq \frac{(-1)^{(k+\nu)}(2\pi a v)^{2k}}{\sqrt{\pi}}\left[-2\pi av\sin(4\pi a v)+(k-\nu^2-1/4)\cos(4\pi a v)\right].
\ee
Using the representation for Bernoulli numbers
\be
\left(1-2^{1-2k}\right)B_{2k}=(-1)^{k+1}\pi\int\limits_0^\infty\frac{dt\,t^{2k}}{\cosh^2(\pi t)},
\ee
we get after performing the summation over $k$,
\bea
I(3,\nu,z,a)=\frac{(-1)^{\nu+1}}{\pi}\hspace{-1mm}\int\limits_0^\infty\hspace{-1mm}\frac{dt}{\cosh^2 t}\left[\cos(4at)\left(\frac{\mu R}{\hbar v_F}\sin(2k_FR)
+\frac{4\nu^2+1}{4}\cos(2k_FR)\right)+2at\sin(4at)\cos(2k_FR)\right].
\eea
Calculating the integrals over $t$, we finally obtain
\be
I(3,\nu,z,a)=(-1)^{\nu+1}\frac{2R^2}{(\hbar v_F)^2}F_1\left[\mu\sin(2k_FR)+\frac{\hbar v_F(4\nu^2-1)}{4R}\cos(2k_FR)+\pi F_2\cos(2k_FR)\right],\quad k_FR\gg1,
\label{I(3,0,z,a)}
\ee
where $F_1$ and $F_2$ are defined in Eq.(\ref{eq:F-definitions-1}). The last expression for $\nu=0,\,1$ leads to the same expressions as were found in graphene
for exchange interactions \cite{klier2015}, while the expression for $\nu=2$ is completely new and corresponds to interaction between impurities on rim sites in considered pseudospin-1 fermion system.

\section{Zero chemical potential and finite temperature}
\label{appendix:zero-chemical-mu}
Asymptotics of the integrals $I_n$ with $n=0,1$ were at least partially analyzed in graphene literature, except the integral $I_2$. However,  in the case of
zero chemical potential, $\mu=0$, such an analysis was not performed to the best of our knowledge. The evaluation of corresponding integrals in the large distance
limit poses a rather complicated task. This is because the leading correction is given by exponentially small term, and thus any power series decomposition can not give the desired result. However, our formula (\ref{Jfinal-withQ}) allows us to analyze the case $\mu=0$ straightforwardly. First, we write the function $Q(s,z=1)$
from Eq.(\ref{Q-z-near1}) in the form
\be
Q(s,1)=-\frac{2\zeta(1+s,1/2)}{(2\pi)^s\cos(\pi s/2)}=-\frac{4}{\pi^{s+1}}\Gamma\left(\frac{1+s}{2}\right)\Gamma\left(\frac{1-s}{2}\right)\sum\limits_{k=0}^\infty\frac{1}{(2k+1)^{s+1}},
\quad {\rm Re}\,s>0.
\ee
Then for the integral (\ref{Jfinal-withQ}) we obtain
\begin{align}
I(\alpha, \nu,1,a)=\frac{2a}{\sqrt{\pi}}\sum\limits_{k=0}^\infty\frac{1}{2\pi i}\int\limits_{\gamma-i\infty}^{\gamma+i\infty} ds\,[\pi a(2k+1)]^{-s-1}\frac{\Gamma\left(\nu+\frac{\alpha+s}{2}\right)\Gamma\left(\frac{\alpha+s}{2}\right)\Gamma\left(\frac{1+s}{2}\right)\Gamma\left(\frac{1-s}{2}\right)}
{\Gamma\left(\frac{\alpha+1+s}{2}\right)\Gamma\left(\nu+1-\frac{\alpha+s}{2}\right)},\quad 0<\gamma<1.
\label{integral-mu=0}
\end{align}
Finally, making the change $s\to 2s-1$ we get the expression in terms of Meijer functions,
\begin{align}
I(\alpha, \nu,1,a)=\frac{4a}{\sqrt{\pi}}\sum\limits_{k=0}^\infty
G^{3,1}_{2,4}\left(\pi^2 a^2(2k+1)^2\Big|\begin{array}{c}0,\frac{\alpha}{2}\\ 0,\frac{\alpha-1}{2},\nu+\frac{\alpha-1}{2},\frac{\alpha-1}{2}-\nu \end{array}\right).
\label{J(alpha,nu,z,a)-final}
\end{align}
The function $G^{31}_{24}(z)$ is an analytic in $z$ function in the sector $|\mbox{arg}z|<\pi$.
To find asymptotic behavior of $J(\alpha, \nu,1,a)$ at large $a$, we use two terms of asymptotic expansion of Meijer function at large argument and then
evaluate the sum. Below we present results for three cases $\nu=0,1,2$:
\begin{align}
\label{eq:asymp-large-a-1}
\hspace{-8mm}I(3,0,1,a)&= -\frac{2 a^2}{\sinh(2\pi a)}\left(\frac{\pi}{\tanh(2\pi a)}-\frac{1}{4a}\right),\quad a>1.  \\
\label{eq:asymp-large-a-2}
\hspace{-8mm}I(3,1,1,a)&=\frac{ 2a^2}{\sinh(2\pi a)}\left(\frac{\pi}{\tanh(2\pi a)}+\frac{3}{4a}\right),\, a>1.\\
\label{eq:asymp-large-a-3}
\hspace{-8mm}I(3,2,1,a)&=\frac{4}{\pi a}-\frac{2a^2}{\sinh(2\pi a)}\left(\frac{\pi}{\tanh(2\pi a)}+\frac{15}{4a}\right),\, a>1.
\end{align}
The last expression contains the power decreasing term $\sim 1/a$ in contrast to the first two expressions. This is because the corresponding
Mellin-Barnes integrand has one pole (at $s=1$) to the right of the integration contour while the integrands for $\alpha=3$ and $\nu=1,2$ do not contain
poles at all in that region. Hence they have only exponentially decreasing terms, for example, the first correction is exponentially small,
$\sim a^2\exp(-2\pi a)$,  at large $a\gg1$. On the other hand, since the expression for $\nu=2$ decreases as $\sim 1/a$ the corresponding integral in Eq.\eqref{eq:In-rewritten} 
has $1/R^4$ decrease with a distance. However, as we find from Eq.\eqref{eq:additional-term-I2} in main text, this power-decreasing term is exactly canceled by the flat-band correction.


\begin{thebibliography}{99}
\bibitem{Ruderman-Kittel}  M.A.Ruderman and C.Kittel, Phys. Rev. {\bf 96}, 99 (1954); T.Kasuya, Prog. Theor. Phys. {\bf 16}, 45 (1956); K. Yosida,
Phys. Rev. {\bf 106}, 893 (1957).
	
\bibitem{Yafet}  Y. Yafet, Phys. Rev. B {\bf 36}, 3948 (1987).
	
\bibitem{Fischer} B. Fischer and M. W. Klein, Phys. Rev. B {\bf 11}, 2025 (1975).
	
\bibitem{BreySarma2007} L. Brey, H. A. Fertig, and S. Das Sarma, Phys. Rev. Lett.  {\bf 99}, 116802 (2007).

\bibitem{saremi2007} S. Saremi, Phys. Rev. B {\bf 76}, 184430 (2007).	

\bibitem{Kogan-doped-2012} E. Kogan, Graphene, {\bf2}, 8 (2013). doi:10.4236/graphene.2013.21002.

\bibitem{sherafati-doped-2011} M. Sherafati and S. Satpathy, Phys. Rev. B {\bf 84}, 125416 (2011).

\bibitem{roslyak2013} O. Roslyak, G. Gumbs, and D. Huang, Journal of Applied Physics {\bf 113}, 123702 (2013).

\bibitem{Cao2019} J. Cao, H.A. Fertig, and Sh. Zhang, Phys. Rev. B {\bf99}, 205430 (2019).

\bibitem{Kogan2019} E. Kogan, C-Journal of Carbon Research {\bf 5}, 14 (2019). doi:10.3390/c5020014.

\bibitem{Kogan2011} E. Kogan, Phys. Rev. B {\bf 84}, 115119 (2011).		

\bibitem{Black2010} A. M. Black-Schaffer, Phys. Rev. B {\bf 81}, 205416 (2010).

\bibitem{sherafati2011} M. Sherafati and S. Satpathy, Phys. Rev. B {\bf 83}, 165425 (2011).
	
\bibitem{klier2015} N. Klier, S. Shallcross, S. Sharma, and O. Pankratov, Phys. Rev. B {\bf 92}, 205414 (2015).

\bibitem{Gorman}P. D.~Gorman, J. M.~Duffy, M. S.~Ferreira, and S. R.~Power, Phys. Rev. B {\bf88}, 085405 (2013).

\bibitem{Parhizgar2013}F.~Parhizgar, M.~Sharafati, R.~Asgari, and S.~Satpathy, Phys. Rev. B {\bf87}, 165429 (2013).

\bibitem{Klier2014}N.~Klier, S.~Shallcross, and O.~Pankratov, Phys. Rev. B {\bf90}, 245118 (2014).

\bibitem{Zera2019}M.~Zare, Phys. Rev. B {\bf100}, 085423 (2019).

\bibitem{Paul2019} G. C. ~Paul, SK Firoz ~Islam, and A.~Saha, Phys. Rev. B {\bf 99}, 155418 (2019).

\bibitem{Kaladzhyan2019}V.~Kaladzhyan, A.A.~Zyuzin, and P.~Simon, Phys. Rev. B {\bf99}, 165302 (2019).

\bibitem{Bradlyn} B. Bradlyn, J. Cano, Zh. Wang, M.G. Vergniory, C. Felser, R.J. Cava, B. Andrei Bernevig, Science, {\bf 353}, aaf5037 (2016).

\bibitem{Sutherland} B. Sutherland, Phys. Rev. B {\bf 34}, 5208 (1986).

\bibitem{Bercioux} D. Bercioux, D. F. Urban, H. Grabert, and W. Hausler, Phys. Rev. A {\bf 80}, 063603 (2009).

\bibitem{Shen2010}R.~Shen, L.B.~Shao, B.~Wang, and D.Y.~Xing, Phys. Rev. B {\bf 81}, 041410(R) (2010).

\bibitem{Green2010}D.~Green, L.~Santos, and C.~Chamon, Phys. Rev. B {\bf82}, 075104 (2010).

\bibitem{Kang2019} M. Kang, L. Ye, S. Fang, J.-S. You, A. Levitan, M. Han, J. I. Facio, C. Jozwiak, A. Bostwick, E. Rotenberg, et. al., Nature Mater. {\bf 19}, 163 (2019).

\bibitem{Slot2017Nature} M. R. Slot, T. S. Gardenier, P. H. Jacobse, G. C. P. van Miert, S. N. Kempkes, S. J. M. Zevenhuizen, C. M. Smith, D. Vanmaekelbergh, and I. Swart, Nature Physics {\bf 13}, 672 (2017). 

\bibitem{Jiang2019Nature} W. Jiang, H. Huang, and F. Liu, Nat. Commun. {\bf 10}, 2207 (2019).

\bibitem{Cui2020Nature} B. Cui, X. Zheng, J. Wang, D. Liu, S. Xie, and B. Huang, Nature Commun. {\bf 11}, 66 (2020).


\bibitem{Malcolm} J. D. Malcolm and E. J. Nicol, Phys. Rev. B {\bf 92}, 035118 (2015).

\bibitem{Lan2012} Z.~Lan, N.~Goldman, and  P. ~Ohberg, Phys. Rev. B {\bf85}, 155451 (2012).

\bibitem{Wang2018}L.~Wang and D.-X.~Yao, Phys. Rev. B {\bf98}, 161403(R) (2018).

\bibitem{Tang2017}P.~Tang, Q.~Zhou, and S.-C.~Zhang, Phys. Rev. Lett. {\bf119}, 206402 (2017).

\bibitem{Raoux} A. Raoux, M. Morigi, J.-N. Fuchs, F. Pi$\acute{e}$chon, and G. Montambaux, Phys. Rev. Lett. {\bf 112}, 026402 (2014).
	
\bibitem{Vidal} J. Vidal, R. Mosseri, and B. Doucot, Phys. Rev. Lett. {\bf 81}, 5888 (1998).

\bibitem{Abilio1999} C. C. Abilio, P. Butaud, Th. Fournier, B. Pannetier, J. Vidal, S. Tedesco, and B. Dalzotto, Phys. Rev. Lett. {\bf 83}, 5102 (1999).

\bibitem{Serret} E. Serret, P. Butaud, and B. Pannetier, Europhys. Lett. {\bf 59}, 225 (2002).

\bibitem{Naud2001} C. Naud, G. Faini, and D. Mailly, Phys. Rev. Lett. {\bf 86}, 5104 (2001).

\bibitem{Rizzi} M. Rizzi, V. Cataudella, and R. Fazio, Phys. Rev. B {\bf 73}, 144511 (2006).	

\bibitem{Malcolm_2014} J.~D.~Malcolm and E.~J.~Nicol, Phys. Rev. B {\bf 90}, 035405 (2014).

\bibitem{Carbotte}E.~Illes, J.~P.~Carbotte, and E.J.~Nicol, Phys. Rev. B {\bf92}, 245410 (2015).

\bibitem{Illes} E.~Illes and E.~J.~Nicol, Phys. Rev. B {\bf94}, 125435 (2016).

\bibitem{Cserti} A.D.~Kovacs, G.~David, B.~Dora, and J.~Cserti, Phys. Rev. B {\bf95}, 035414 (2017).

\bibitem{Biswas}T.~Biswas and T.~K.~Ghosh, J. Phys.: Condens. Matter, {\bf28}, 495302 (2016).

\bibitem{Xu}Y.~Xu and L.-M. Duan, Phys. Rev. B {\bf96}, 155301 (2017).

\bibitem{Islam}SK Firoz Islam and P. Dutta, Phys. Rev. B {\bf 96}, 045418 (2017).

\bibitem{Biswas2018} T.~Biswas and T.K.~Ghosh, J. Phys.: Condens. Matter {\bf 30}, 075301 (2018)

\bibitem{Firoz_Islam} M.-W. Alam, B. Souayeh, SK F. Islam, J. Phys.: Condens. Matter {\bf31}, 485303 (2019).

\bibitem{Oriekhov2018LTP} D.O. Oriekhov, E.V. Gorbar, and V.P. Gusynin, Low Temperature Physics {\bf 44}, 1313 (2018).

\bibitem{Coulomb_alphaT3} E. V. Gorbar, V. P. Gusynin, and D. O. Oriekhov, Phys. Rev. B {\bf 99}, 155124 (2019).

\bibitem{Bugajko2019} O. V. Bugaiko and D. O. Oriekhov, J. Phys.: Condens. Matter {\bf 31} 325501 (2019).

\bibitem{Cao-twisted} Y. Cao, V. Fatemi, Sh. Fang, K. Watanabe, T. Taniguchi, E. Kaxiras and P. Jarillo-Herrero, Nature {\bf 556}, 43 (2018).

\bibitem{Khodel2017}V.A.~Khodel, J. Low Temp. Phys. {\bf191}, 14 (2017).

\bibitem{Leykam2018}D.~Leykam, A.~Andreanov, and S.~Flach, Adv. Phys. X {\bf3}, 1473052 (2018).

\bibitem{Delplace2017} P. Delplace, J. B. Marston, and A. Venaille,  Science {\bf 358}, 1075 (2017).

\bibitem{Hasan2018Review}  H. Zheng, and M. Zahid Hasan, Adv. Phys. X, {\bf 3}:1, 1466661 (2018).

\bibitem{PrudnikovII} A.P.~Prudnikov, Yu.A.~Brychkov, and O.I.~Marichev, Integrals and Series. Special
functions. V.II, Nauka, Moskow, 1983.

\bibitem{Prudnikov3} A.P.~Prudnikov, Yu.A.~Brychkov, and O.I.~Marichev, Integrals and Series. Special
functions. V.III, Nauka, Moskow, 1983.

\bibitem{Whittaker} E. T. Whittaker and Watson, A Course of Modern Analysis (4th edition), Cambridge University Press, Cambridge, 1927.

\bibitem{Bateman1} H.~Bateman and A.~Erdelyi, Higher Transcendental Functions, V.1, MC Graw-Hill Book Co.,New York, 1953.

\bibitem{Drukarev} Yu. N. Demkov and G.F. Drukarev, Sov. Phys. JETP {\bf22}, 182 (1965).
	
\end{thebibliography}
\end{document}